\documentclass[sn-standardnature]{sn-jnl}

\usepackage{caption}

\jyear{2023}%

\theoremstyle{thmstyleone}%
\theoremstyle{thmstyletwo}%

\theoremstyle{thmstylethree}%

\begin{document}

\title[ ]{Itinerant and topological excitations in a honeycomb spiral spin liquid candidate}


\author[1]{Yuqian Zhao}\equalcont{These authors contributed equally to this work.}
\author[2]{Xuping Yao}\equalcont{These authors contributed equally to this work.}
\author[1]{Xun Chen}
\author[1]{Zongtang Wan}
\author[1]{Zhaohua Ma}
\author[3]{Xiaochen Hong}
\author*[1]{Yuesheng Li}\email{yuesheng\_li@hust.edu.cn}

\affil[1]{\orgdiv{Wuhan National High Magnetic Field Center and School of Physics}, \orgname{Huazhong University of Science and Technology}, \orgaddress{\city{Wuhan}, \postcode{430074}, \country{China}}}
\affil[2]{\orgdiv{Kavli Institute for Theoretical Sciences}, \orgname{University of Chinese Academy of Sciences}, \orgaddress{\city{Beijing}, \postcode{100190}, \country{China}}}
\affil[3]{\orgdiv{Department of Applied Physics and Center of Quantum Materials and Devices}, \orgname{Chongqing University}, \orgaddress{\city{Chongqing}, \postcode{401331}, \country{China}}}

\maketitle


The frustrated insulating magnet can stabilize a spiral spin liquid, arising from cooperative fluctuations among a subextensively degenerate manifold of spiral configurations, with ground-state wave vectors forming a continuous contour or surface in reciprocal space. The atomic-mixing-free honeycomb antiferromagnet GdZnPO has recently emerged as a promising spiral spin-liquid candidate, hosting nontrivial topological excitations. Despite growing interest, the transport and topological properties of spiral spin liquids remain largely unexplored experimentally. Here, we report transport measurements on high-quality, electrically insulating GdZnPO single crystals. We observe a giant low-temperature magnetic thermal conductivity down to $\sim$50 mK, described by $\kappa_{xx}^\mathrm{m}$ $\sim$ $\kappa_0+\kappa_1T$, where both $\kappa_0$ and $\kappa_1$ are positive constants associated with excitations along and off the spiral contour in reciprocal space, respectively. This behavior parallels the magnetic specific heat, underscoring the presence of mobile low-energy excitations intrinsic to the putative spiral spin liquid. Furthermore, the observed positive thermal Hall effect confirms the topological nature of at least some of these excitations. Our findings provide key insights into the itinerant and topological properties of low-lying spin excitations in the spiral spin-liquid candidate.

\bigskip

\noindent
In strongly correlated magnets, frustration can give rise to exotic phases such as spin liquids, characterized by fractionalized excitations, long-range entanglements, and topological order~\cite{savary2016quantum,zhou2017quantum}. These phases hold potential for applications in topological quantum computation~\cite{nayak2008non}, spintronics devices~\cite{jungwirth2016antiferromagnetic,gao2020fractional}, and understanding high-temperature superconductivity~\cite{RevModPhys.78.17}. Notably, some frustrated magnets with large spin quantum numbers ($S$) can stabilize spiral spin liquids (SSLs). The SSL emerges from cooperative fluctuations among subextensively degenerate spiral configurations, with ground-state wave vectors ($\mathbf{Q}_\mathrm{G}$) forming a continuous contour or surface in reciprocal space for two- (2D) or three-dimensional (3D) systems~\cite{bergman2007order,yao2021generic}. For instance, the SSL is predicted to stabilize in the easy-plane ($D$ $\geq$ 0) frustrated honeycomb-lattice model: $\mathcal{H} = J_1\sum_{\langle j0,j1\rangle}\mathbf{S}_{j0}\cdot\mathbf{S}_{j1}+J_2\sum_{\langle\langle j0,j2\rangle\rangle}\mathbf{S}_{j0}\cdot\mathbf{S}_{j2}+D\sum_{j0}(S_{j0}^z)^2-\mu_0Hg\mu_\mathrm{B}\sum_{j0}S_{j0}^z$, where $J_1$ and $J_2$ are the first- and second-nearest-neighbor Heisenberg couplings, $H$ is the applied magnetic field along the $z$ axis, and $g$ is the $g$ factor~\cite{PhysRevB.81.214419,owerre2017topological,PhysRevB.106.035113,okumura2010novel,PhysRevB.100.224404,PhysRevResearch.4.013121}. This SSL is predicted to remain stable down to very low temperatures~\cite{PhysRevResearch.4.013121}, with $\mathbf{Q}_\mathrm{G}$ forming a continuous contour around the $\Gamma\{0, 0\}$ point for 1/2 $>$ $\mid$$J_2$/$J_1$$\mid$ $>$ 1/6, or around the K$\{1/3, 1/3\}$ point for $\mid$$J_2$/$J_1$$\mid$ $>$ 1/2.

Recently, the honeycomb antiferromagnet GdZnPO emerged as a promising candidate for realizing the above prototypical model experimentally (see Fig.~\ref{fig1}), with $S$ = 7/2, $J_1$ $\sim$ $-$0.39 K, $J_2$ $\sim$ 0.57 K, $D$ $\sim$ 0.30 K, and $g$ $\sim$ 2~\cite{PhysRevLett.133.236704}. With $J_2/J_1$ $\sim$ $-$1.5, $\mathbf{Q}_\mathrm{G}$ is expected to form a degenerate spiral contour around the K point below the crossover temperature $T^*$ $\sim$ 2 K, accompanied by low-energy topological excitations on the sublattices~\cite{PhysRevLett.133.236704}. These excitations, including spin and momentum vortices, offer potential for applications in antiferromagnetic spintronics without magnetic field leakage~\cite{jungwirth2016antiferromagnetic,gao2020fractional}, topologically protected memory and logic operations~\cite{yao2013topologically}, fracton gauge theory~\cite{PhysRevB.95.115139,nandkishore2019fractons,pretko2020fracton}, and beyond.
Moreover, the spiral contour's approximate U(1) symmetry in momentum space~\cite{PhysRevLett.133.236704,PhysRevResearch.4.023175} may enrich symmetry-breaking states, such as unidirectional spin-density waves~\cite{PhysRevB.108.224423}, making SSLs promising platforms for exploring related phenomena like unidirectional pair-density waves. However, the transport and topological properties of SSLs remain largely unexplored experimentally.

The specific heat ($C_\mathrm{m}$) of the generic 2D SSL is unusually large, following $C_\mathrm{m}$ $\sim$ $C_0$+$C_1T$ at low temperatures, within the spherical approximation~\cite{bergman2007order,yao2021generic}. In the SSL, $\mathbf{Q}_\mathrm{G}$ fluctuates along the continuous contour in reciprocal space, in sharp contrast to conventional magnetic states where $\mathbf{Q}_\mathrm{G}$ adopts discrete values. Analogous to an ideal gas, this implies the presence of low-temperature spin degrees of freedom along the spiral contour, giving rise to a finite residual specific heat $C_0$. Therefore, $C_0$ arises from zero-energy excitations along the degenerate continuous contour in the classical limit ($S$ $\to$ $\infty$), while the $C_1T$ term reflects low-energy excitations off the contour. To our knowledge, this low-$T$ behavior of $C_\mathrm{m}$ $\sim$ $C_0$+$C_1T$ had not been experimentally observed in any other compounds until our previous report on the SSL candidate GdZnPO~\cite{PhysRevLett.133.236704}. In GdZnPO, measurements reveal $C_0$ = 0.9-1.4 JK$^{-1}$/mol and $C_1$ = 1.6-3.4 JK$^{-2}$/mol below the crossover field $\mu_0H_\mathrm{c}$ $\sim$ 12 T, which separates the putative SSL and the polarized phase and is analytically given by $\mu_0H_\mathrm{c}$ = $S[2D+3J_1+9J_2+J_1^2/(4J_2)]/(g\mu_\mathrm{B})$ in the classical limit (see Fig.~\ref{fig1}c). These results support the stability of the honeycomb SSL down to at least $\sim$50 mK. For $H$ $\geq$ $H_\mathrm{c}$, the GdZnPO spin system becomes nearly fully polarized in a nondegenerate ferromagnetic phase. The field-induced transition from this ferromagnetic state ($H$ $\geq$ $H_\mathrm{c}$) to the degenerate SSL ($H$ $<$ $H_\mathrm{c}$) results in a significant entropy increase, decalescence, and a giant magnetocaloric effect--exceeding other known materials--underscoring GdZnPO's potential for magnetic cooling applications down to $\sim$36 mK~\cite{zhao2025giant}.

Despite extensive efforts~\cite{PhysRevLett.120.067202,PhysRevB.107.184423,hong2024phonon,PhysRevLett.117.267202,rao2021survival,10.1126/science.1188200,PhysRevX.9.041051,PhysRevLett.123.247204,PhysRevB.101.140407,shimozawa2017quantum,li2020possible,huang2022thermal,PhysRevB.104.104403,shen2016quantum,PhysRevLett.127.267202,PhysRevLett.121.097203,PhysRevB.106.L220406,jeon2024one,tokiwa2016possible,PhysRevLett.131.256701,PhysRevLett.129.167201,PhysRevResearch.2.013099,PhysRevB.96.081111}, magnetic insulators with large thermal conductivity in the low-temperature limit remain exceptionally rare. GdZnPO single crystals are transparent insulators with no evident atomic-mixing disorder~\cite{nientiedt1998equiatomic,lincke2008magnetic,PhysRevLett.133.236704}. At low temperatures, the observed giant magnetic specific heat ($C_\mathrm{m}$)~\cite{PhysRevLett.133.236704} reflects a high density of low-energy spin excitations, which can lead to significant magnetic thermal conductivity $\kappa_{xx}^\mathrm{m}$ $\sim$ $\frac{C_\mathrm{m}\lambda_\mathrm{m}v_\mathrm{m}}{3N_\mathrm{A}V_0}$, provided the mean free path ($\lambda_\mathrm{m}$) and velocity ($v_\mathrm{m}$) remain nonzero. Here, $V_0$ = 67.6 \AA$^3$ represents the volume per formula. Consequently, GdZnPO may serve as a rare example of an insulator that supports mobile unconventional spin excitations at low temperatures, offering an excellent platform for investigating the transport and topological properties of a 2D SSL candidate.

In this work, we conducted comprehensive transport measurements on five independent GdZnPO single-crystal samples. The sample with higher crystal quality exhibited reduced electric conductivity but enhanced thermal conductivity, indicating a nonzero intrinsic $\kappa_{xx}^\mathrm{m}$ in the low-temperature limit. At low temperatures, the behavior $\kappa_{xx}^\mathrm{m}$ $\sim$ $\kappa_1T$+$\kappa_0$ is clearly observed in the highest-quality crystals. This behavior aligns well with the giant magnetic specific heat, $C_\mathrm{m}$ $\sim$ $C_1T$+$C_0$, suggesting the presence of mobile, high-density spin excitations in GdZnPO. Additionally, a positive thermal Hall effect is observed, supporting the emergence of nonzero Chern numbers. Our findings suggest the itinerant and topological nature of low-lying excitations in the putative SSL, motivating further study.

\bigskip

\section*{Results}

\noindent {\bf Toward intrinsic transport properties}

\begin{figure}[H]
\begin{center}
\includegraphics[width=9cm,angle=0]{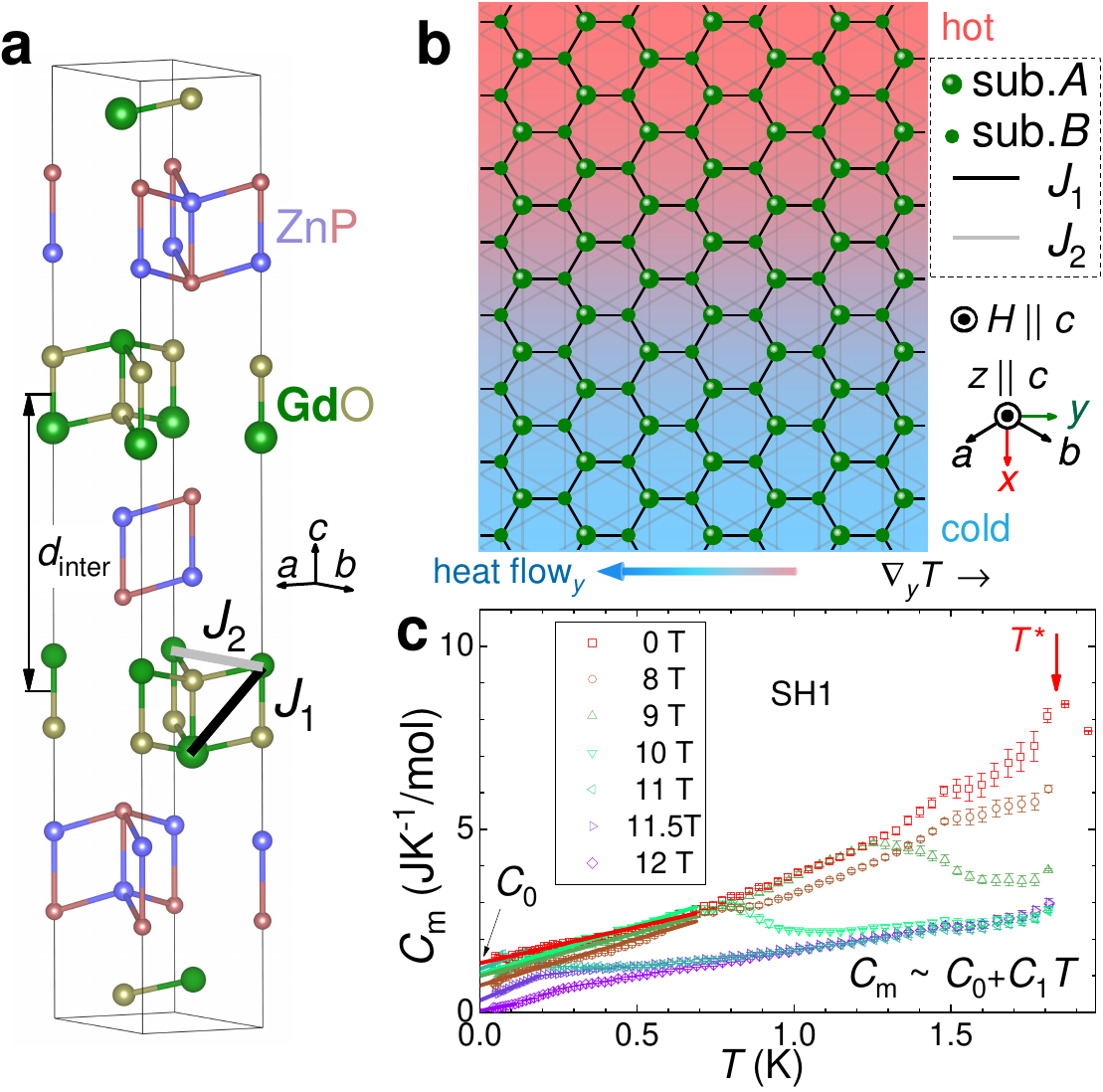}
\captionsetup{labelformat=empty,font={small}}
\caption{\textbf{Fig. 1 $\vert$ Lattice structure and magnetic specific heat of GdZnPO.}
\textbf{a}, Crystal structure showing magnetic GdO layers separated by nonmagnetic ZnP layers, with an average interlayer distance $d_\mathrm{inter}$ = $c$/3 ($\sim$10.2 \AA). The first- and second-nearest-neighbor exchanges $J_1$ and $J_2$ are indicated. Thin lines denote the unit cell. \textbf{b}, Honeycomb lattice of Gd$^{3+}$ ions. Inset: coordinate system for spin components and thermal transport. \textbf{c}, Magnetic specific heat $C_\mathrm{m}$ of sample SH1 in selected magnetic fields ($\mu_0H$) applied along the $c$ axis. Colored lines are linear fits $C_\mathrm{m}$ $\sim$ $C_0$+$C_1T$ below the temperature where $C_\mathrm{m}$ begins to exhibit linear behavior: below 0.7 K for $\mu_0H$ $<$ 11 T, below 0.15 K at $\mu_0H$ = 11 T, below 0.2 K at $\mu_0H$ = 11.5 T, and below 0.12 K at $\mu_0H$ = 12 T. The fitted parameters $C_0$ and $C_1$ are shown in Fig.~\ref{fig3}f. Crossover temperature $T^*$ is marked, and error bars, 1$\sigma$ s.e.}
\label{fig1}
\end{center}
\end{figure}

\noindent Figure~\ref{fig1}a shows the crystal structure of GdZnPO~\cite{nientiedt1998equiatomic,lincke2008magnetic,PhysRevLett.133.236704}. The first-nearest-neighbor coupling, $J_1$ (with $\mid$Gd-Gd$\mid_1$ $\sim$ 3.7 \AA), and the second-nearest-neighbor coupling, $J_2$ (with $\mid$Gd-Gd$\mid_2$ = $a$ $\sim$ 3.9 \AA), are mediated by Gd-O-Gd exchanges within the magnetic GdO layers. These magnetic layers are well separated by nonmagnetic ZnP layers, with an average interlayer distance of $d_\mathrm{inter}$ $\sim$ 10.2 \AA, indicating negligible interlayer coupling. The magnetic Gd$^{3+}$ ions form a $J_1$-$J_2$ frustrated quasi-2D honeycomb lattice (Fig.~\ref{fig1}b). Because of the zero orbital quantum number ($L$ = 0) and negligible spin-orbit coupling of Gd$^{3+}$, interaction anisotropy, such as Dzyaloshinsky-Moriya (DM) anisotropy, is expected to be minimal. This is confirmed by the measured $g$ = 2.01(2)~\cite{PhysRevLett.133.236704}, closely matching the free electron value ($g_\mathrm{e}$ = 2.0023). The maximum DM interaction strength is estimated to be about ($\mid$$g-g_\mathrm{e}$$\mid$/$g_\mathrm{e}$)$J_2$ $\sim$ 1.4\%$J_2$ ($\sim$0.01 K)~\cite{PhysRev.120.91} in GdZnPO.

\begin{figure}[h]
\begin{center}
\includegraphics[width=13.4cm,angle=0]{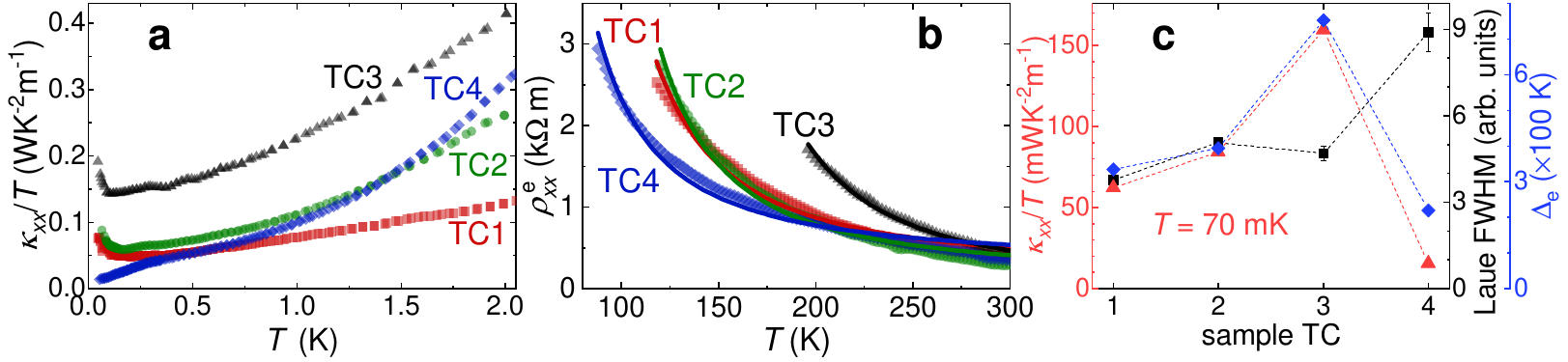}
\captionsetup{labelformat=empty,font={small}}
\caption{\textbf{Fig. 2 $\vert$ Sample dependence of transport properties measured on GdZnPO crystals.}
\textbf{a}, Zero-field thermal conductivity ($\kappa_{xx}$) of four samples, plotted as $\kappa_{xx}$/$T$ vs $T$, respectively. \textbf{b}, Zero-field electrical resistivity ($\rho_{xx}^\mathrm{e}$). Colored lines show Arrhenius-law fits, $\rho_{xx}^\mathrm{e}$ = $\rho_\infty^\mathrm{e}\exp$($\Delta_\mathrm{e}$/$T$), with fitted gap values $\Delta_\mathrm{e}$ shown in \textbf{c}. \textbf{c}, Sample dependence of $\kappa_{xx}$/$T$ at 70 mK, average Laue x-ray diffraction full width at half maximum (Laue FWHM), and $\Delta_\mathrm{e}$. In \textbf{c}, error bars, 1$\sigma$ s.e.}
\label{fig2}
\end{center}
\end{figure}

Because thermal conductivity is highly sensitive to crystal quality, we first measured the low-temperature thermal conductivity of four randomly selected crystals (Supplementary Note 1 and Supplementary Figs. 1-6), followed by Laue x-ray diffraction (XRD) at 300 K and electrical resistivity below 300 K, to identify the intrinsic transport properties of GdZnPO. Figure~\ref{fig2}a shows the zero-field longitudinal thermal conductivity ($\kappa_{xx}$) data for samples TC1-TC4. The measured $\kappa_{xx}$/$T$ vs. $T$ curves are broadly similar across all investigated samples. However, sample TC4 exhibits a downturn below $\sim$0.3 K, contrasting with the behavior of the other samples. Such low-temperature suppression of $\kappa_{xx}$/$T$ has also been reported in various magnetic systems, including PbCuTe$_2$O$_6$~\cite{PhysRevLett.131.256701}, YbMgGaO$_4$~\cite{rao2021survival}, Na$_2$Co$_2$TeO$_6$~\cite{hong2024phonon}, and Cu(C$_6$H$_5$COO)$_2\cdot$3H$_2$O~\cite{PhysRevLett.129.167201}. This phenomenon has been attributed to spin gap opening, many-body localization, and/or disorder effects that suppress the transport of spin excitations~\cite{PhysRevLett.131.256701,rao2021survival,hong2024phonon,PhysRevLett.129.167201}. Laue XRD measurements were conducted to assess the crystallinity of each sample. Sample TC4 exhibits significantly broader reflections than the other samples, as shown in Fig.~\ref{fig2}c and the raw Laue XRD patterns in Supplementary Fig. 3, indicating a smaller average grain size. Crystal grain boundaries and other structural imperfections likely scatter quasi-particles, reducing their mean free path and thereby suppressing the thermal conductivity at low temperatures.

The electric resistivity ($\rho_{xx}^\mathrm{e}$) data measured for the four samples are presented in Fig.~\ref{fig2}b. As the temperature decreases below 300 K, $\rho_{xx}^\mathrm{e}$ ($\geq$ 300 $\Omega$m) increases, approximately following the Arrhenius law, $\rho_{xx}^\mathrm{e}$ = $\rho_\infty^\mathrm{e}\exp$($\Delta_\mathrm{e}$/$T$), where $\rho_\infty^\mathrm{e}$ and $\Delta_\mathrm{e}$ denote the high-temperature resistivity limit and the thermally activated gap, respectively. Below $\sim$90 K (for sample TC4) to 200 K (for sample TC3), the resistances ($\sim$0.7-1.4 M$\Omega$ at 300 K) exceeded the measurement range ($\leq$6 M$\Omega$) of the physical property measurement system (PPMS). The gap $\Delta_\mathrm{e}$ is roughly proportional to the low-temperature $\kappa_{xx}$/$T$, while inversely correlated with the full width at half maximum (FWHM) of Laue XRD reflections (see Fig.~\ref{fig2}c). In insulating materials, crystal imperfections can introduce charge carriers and impair electrical insulation~\cite{PhysRevResearch.2.013099,PhysRevB.86.045314}.

Consequently, higher-quality GdZnPO crystals with sharper Laue XRD reflections exhibit improved electrical insulation (with larger $\Delta_\mathrm{e}$) and enhanced low-temperature thermal conductivity. A similar trend has also been reported in the electrically insulating magnet 1$T$-TaS$_{2}$~\cite{PhysRevResearch.2.013099}. These observations firmly establish the intrinsic low-temperature properties of GdZnPO: (1) electrical insulation and (2) the presence of itinerant magnetic excitations. Notably, the high-quality sample TC3 shows a large $\kappa_{xx}$/$T$ value of $\sim$160 mWK$^{-2}$m$^{-1}$ at 70 mK and 0 T (the phonon contribution $\leq$ 6 mWK$^{-2}$m$^{-1}$ is negligible, see below), exceeding that of most other known electrically insulating magnets.

At low temperatures below 1 K, most of the thermal conductivity data across various measured fields can be well described by $\kappa_{xx}$ $\sim$ $\kappa_0$+$\kappa_1T$+$K_\mathrm{p}T^3$ (Supplementary Fig. 11). Here, the phonon prefactor $K_\mathrm{p}$ is given by $K_\mathrm{p}$ $\sim$ $\frac{4\pi^4k_\mathrm{B}\lambda_\mathrm{p}v_\mathrm{p}}{5ZV_0\Theta_\mathrm{D}^3}$, where $v_\mathrm{p}$ = $\frac{k_\mathrm{B}\Theta_\mathrm{D}}{\hbar}(\frac{ZV_0}{6\pi^2})^{\frac{1}{3}}$ ($\sim$2,900 m/s) represents the phonon mean velocity, $\Theta_\mathrm{D}$ $\sim$ 117.5 K is the Debye temperature determined from the specific heat of the nonmagnetic reference compound YZnPO (Supplementary Note 2 and Supplementary Fig. 16)~\cite{PhysRevB.105.024418,Liu2021Frustrated}, $Z$ = 6 is the number of formulas per unit cell, and $\lambda_\mathrm{p}$ is the phonon mean free path. From the fitted values of $K_\mathrm{p}$ = 0.023, 0.052, 0.069, and 0.066 WK$^{-4}$m$^{-1}$, we obtain $\lambda_\mathrm{p}$ $\sim$ 4.8, 10.9, 14.5, and 13.8 $\mu$m for samples TC1-TC4, respectively---much smaller than the corresponding average sample width $W$ = 2$\sqrt{A/\pi}$ ($\sim$251, 353, 252, and 258 $\mu$m), where $A$ denotes the cross-sectional area. Assuming $\lambda_\mathrm{p}$ = $W$, the maximum phonon contribution to $\kappa_{xx}$/$T$ is calculated as $K_\mathrm{p}T^2$ $\sim$ 6 mWK$^{-2}$m$^{-1}$ at $T$ = 70 mK for sample TC3.

In practice, fitting low-$T$ thermal conductivity data often yields a $\lambda_\mathrm{p}$ that is significantly smaller than $W$~\cite{PhysRevLett.131.256701,gofryk2014anisotropic,PhysRevB.104.144426,li2020possible}, as summarized in Supplementary Tab. 1. To our knowledge, there is no widely accepted explanation for this discrepancy. A natural speculation is that it arises from phonon scattering at grain boundaries or other imperfections that limit the phonon mean free path, despite the macroscopic sample dimensions. However, other studies---including our previous work on high-quality single crystals of $\alpha$-CoV$_2$O$_6$ using a similar experimental setup~\cite{zhao2024quantum}---have reported a $\lambda_\mathrm{p}$ comparable to $W$~\cite{PhysRevLett.110.217209}. These suggest that the crystallinity of the randomly selected samples TC1-TC4 may simply have been insufficient. In the next section, we identify a higher-quality GdZnPO crystal, TC5, which exhibits sharp Laue photographs, improved electrical insulation, and significantly enhanced low-temperature thermal conductivity.

The average grain size may be roughly estimated from $\lambda_\mathrm{p}$ ($\geq$4.8 $\mu$m), which is usually smaller than $W$, yet still more than four orders of magnitude larger than the lattice constant $a$ $\sim$ 3.9~\AA. Theoretically, finite-size effects in thermodynamic properties become negligible when considering large spin clusters---on the order of 10$^4\times$10$^4\times$10$^4$ sites. Therefore, no significant sample dependence is expected for the thermodynamic properties of our GdZnPO crystals. We have measured the low-temperature specific heat on two additional samples (SH1 and SH2) with different crystallinity qualities (Supplementary Fig. 9) to examine possible sample dependence. The specific heat shows negligible sample dependence, with $\mid$$C_\mathrm{m}$(SH2)/$C_\mathrm{m}$(SH1)$-1$$\mid$ $\leq$ 0.1 at 0 T, which falls well within the error margins of two independent ultra-low-temperature measurements. In contrast, thermal transport exhibits evident sample dependence---likely due to scattering from grain boundaries or other structural imperfections. For example, $\kappa_{xx}$(TC5)/$\kappa_{xx}$(TC4)$-$1 $\sim$ 20-350 (Supplementary Fig. 10). This is reasonable given that $\kappa_{xx}$ $\sim$ $\frac{C\lambda v}{3N_\mathrm{A}V_0}$, where $\lambda$ is highly sensitive to sample crystallinity. Very similar sample dependence in both specific heat and thermal conductivity had been reported in several other frustrated magnets, including YbMgGaO$_4$~\cite{li2015gapless,paddison2017continuous, PhysRevLett.117.267202, rao2021survival} and Na$_2$BaCo(PO$_4$)$_2$~\cite{li2020possible,huang2022thermal}.

\bigskip

\noindent {\bf Thermal conductivity of the highest-quality crystal}

\begin{figure}[h]
\includegraphics[width=13.4cm,angle=0]{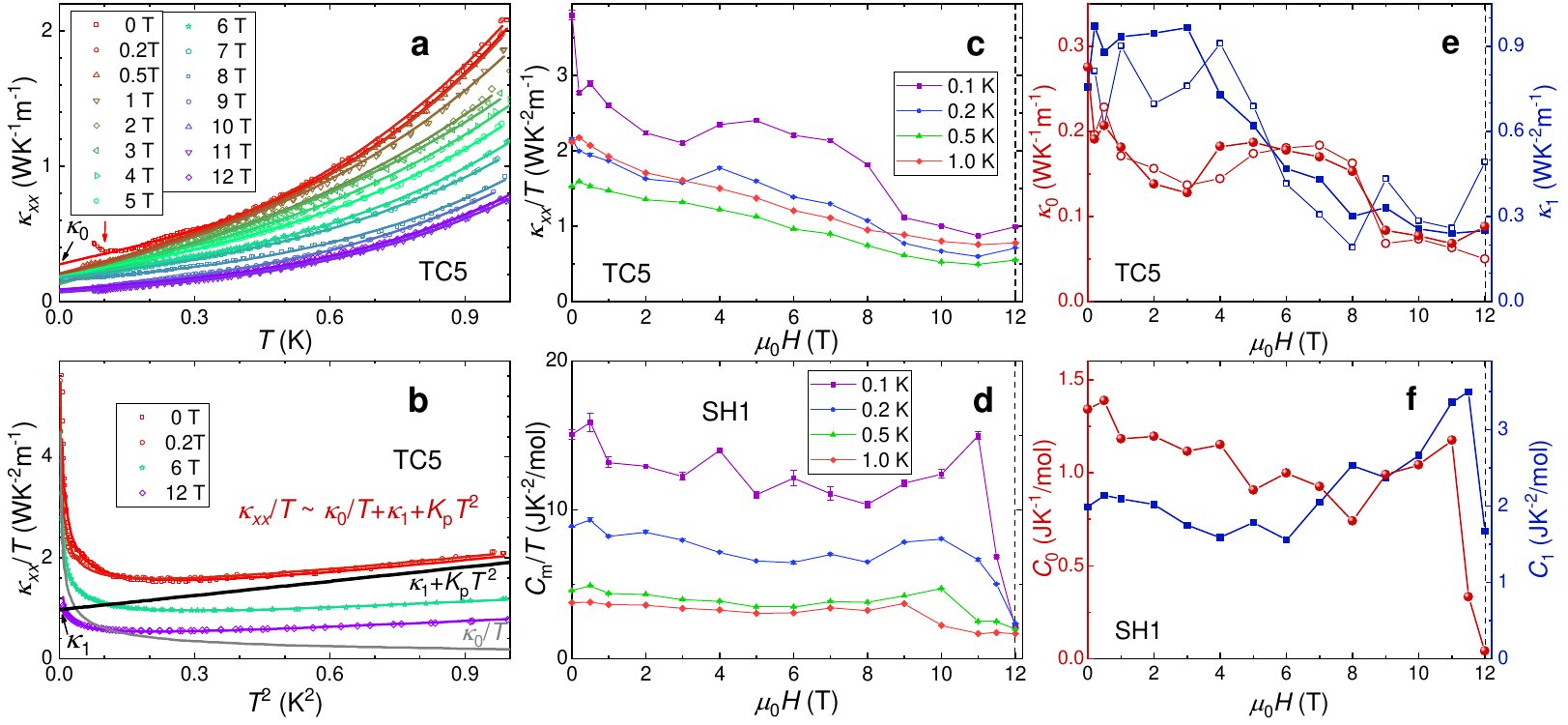}
\captionsetup{labelformat=empty,font={small}}
\caption{\textbf{Fig. 3 $\vert$ Thermal conductivity and specific heat of GdZnPO with magnetic field along the $c$ axis.}
\textbf{a}, Thermal conductivity $\kappa_{xx}$ of sample TC5 at various fields, with fits $\kappa_{xx}$ $\sim$ $\kappa_0+\kappa_1T+K_\mathrm{p}T^3$ below 1 K (colored lines). Fitted parameters $\kappa_0$ and $\kappa_1$ are shown in \textbf{e} (solid symbols). The vertical arrow marks the upturn onset in zero-field $\kappa_{xx}$ as $T$ decreases. \textbf{b}, $\kappa_{xx}$/$T$ vs $T^2$ for selected data. The grey and black lines represent the $\kappa_0$/$T$ and $\kappa_1+K_\mathrm{p}T^2$ components, respectively, of the 0.2 T fit. \textbf{c},\textbf{d}, Field dependence of $\kappa_{xx}$/$T$ and magnetic specific heat $C_\mathrm{m}$/$T$ at selected temperatures (error bars: 1$\sigma$ s.e.). \textbf{e},\textbf{f}, Field dependence of fitted parameters $\kappa_0$, $\kappa_1$, $C_0$, and $C_1$. In \textbf{e}, open symbols show the fitted values of $\kappa_0$ and $\kappa_1$ from linear fits $\kappa_{xx}$ $\sim$ $\kappa_0+\kappa_1T$ at $T$ $\leq$ 0.14 K, where the $K_\mathrm{p}T^3$ term ($\leq$ 0.018 WK$^{-1}$m$^{-1}$) is negligible. Dashed lines in \textbf{c}-\textbf{f} mark the crossover field $\mu_0H_\mathrm{c}$ $\sim$ 12 T.}
\label{fig3}
\end{figure}

\noindent The thermal conductivity data of sample TC5 are expected to reflect the intrinsic properties of GdZnPO for the following reasons: (1) Sample TC5 exhibit sharp Laue photographs (Supplementary Fig. 9), and a room-temperature resistivity of $\rho_{xx}^\mathrm{e}$(300 K) $\sim$ 11 k$\Omega$m over 26 times higher than those of TC1-TC4 ($\sim$0.30-0.42 k$\Omega$m, see Fig.~\ref{fig2}b). (2) In the case of GdZnPO, the low-temperature drop in $\kappa_{xx}$/$T$ is mainly observed in sample TC4 (see Fig.~\ref{fig2}a), which exhibits broader Laue patterns and poorer electrical insulation (see above), suggesting that the suppression is caused by disorder effects. In contrast, sample TC5 shows no such drop. Instead, its thermal conductivity is well described by $\kappa_{xx}$ $\sim$ $\kappa_0+\kappa_1T+K_\mathrm{p}T^3$ below 1 K, across various magnetic fields up to 12 T (see Fig.~\ref{fig3}a,b). Remarkably, in zero field, sample TC5 even exhibits an unusual upturn in $\kappa_{xx}$ below $\sim$0.1 K, which is highly reproducible. This upturn is suppressed by a small magnetic field of $\sim$0.2 T. A similar phenomenon was previously reported in the Kitaev material Na$_2$Co$_2$TeO$_6$, where the low-temperature rise in zero-field $\kappa_{xx}$/$T$ was attributed to the recovery of thermal transport by itinerant excitations at ultra low temperatures~\cite{PhysRevB.107.184423}. (3) From the fits, we obtained a significantly enhanced $K_\mathrm{p}$ $\sim$ 0.94 WK$^{-4}$m$^{-1}$ (Fig.~\ref{fig3}b), and $\lambda_\mathrm{p}$ $\sim$ 0.20 mm, which closely matches $W$ = 0.21 mm, for sample TC5. (4) As shown in Supplementary Fig. 12d, the fitted $\kappa_0$ first suddenly increases from nearly zero at 0 T to a large value at 0.5-0.6 T in samples TC1-TC3. Since $C_0$ is already significant at 0 T (see Fig.~\ref{fig3}f) due to the putative SSL contour, we attribute this phenomenon to a disorder effect: the excitations along the spiral contour may be scattered by abnormal spin configurations near the grain boundaries, and a weak applied field of $\sim$0.5 T may suppress this scattering caused by local perturbations from crystal imperfections. In the highest-quality sample TC5, this disorder effect is absent (see Fig.~\ref{fig3}e), and the field dependence of $\kappa_0$ resembles that of $C_0$ (see Fig.~\ref{fig3}f), expect for the sudden drop of $C_0$ near the crossover field of $\sim$12 T, which is discussed later.

The magnetic thermal conductivity, $\kappa_{xx}^\mathrm{m}$ $\sim$ $\kappa_0+\kappa_1T$, is also significantly enhanced in the highest-quality sample TC5, with $\kappa_0$ $\sim$ 0.05-0.28 WK$^{-1}$m$^{-1}$ and $\kappa_1$ $\sim$ 0.25-0.97 WK$^{-2}$m$^{-1}$ across 0-12 T (see Fig.~\ref{fig3}e and Supplementary Fig. 12). In conventionally condensed states, the specific heat ($C$) and thermal conductivity ($\kappa_{xx}$) of (quasi-)particles following Fermi-Dirac or Bose-Einstein statistics exhibit a low-temperature dependence of $\propto$ $T^\beta$ or $\propto$ $\exp(-\Delta/T)$, where $\beta$ $\geq$ 1 and $\Delta$ $>$ 0. Examples include $\beta = 1$ for fermionic systems, $\beta = 2$ for 2D antiferromagnetic bosonic systems, and $\Delta$ $>$ 0 for gapped systems. This behavior leads to either a finite constant or a vanishing $C/T$ and $\kappa_{xx}/T$ as $T \to 0$ K. Surprisingly, $\kappa_{xx}$/$T$ measured on the highest-quality GdZnPO crystal TC5 exhibits a robust upturn, following $\kappa_{xx}$/$T$ $\sim$ $\kappa_0$/$T$$+\kappa_1+K_\mathrm{p}T^2$ below $\sim$1 K (see Fig.~\ref{fig3}b). Across fields up to 12 T, the magnetic thermal conductivity, $\kappa_{xx}^\mathrm{m}$ $\sim$ $\kappa_0+\kappa_1T$, aligns with the magnetic specific heat behavior, $C_\mathrm{m}$ $\sim$ $C_1T$+$C_0$ (Fig.~\ref{fig1}c), consistent with the relation $\kappa_{xx}^\mathrm{m}\propto C_\mathrm{m}$. For excitations along and off the putative SSL contour, we obtained $\lambda_\mathrm{m0}v_\mathrm{m0}$ $\sim$ $3N_\mathrm{A}V_0\kappa_0$/$C_0$ $\sim$ 7.1-150 mm$^2$s$^{-1}$, and $\lambda_\mathrm{m1}v_\mathrm{m1}$ $\sim$ $3N_\mathrm{A}V_0\kappa_1$/$C_1$ $\sim$ 8.8-70 mm$^2$s$^{-1}$ for sample TC5, depending on the applied magnetic field (see Supplementary Fig. 13)---values significantly smaller than the phonon parameter $\lambda_\mathrm{p}v_\mathrm{p}$ $\sim$ 0.57 m$^2$s$^{-1}$. Using the mean magnon velocity calculated from spin-wave theory for spin excitations off the SSL contour, $v_\mathrm{m1}$ $\sim$ 110 ms$^{-1}$ (Supplementary Note 3 and Supplementary Figs. 9 and 10), we estimated $\lambda_\mathrm{m1}$ $\sim$ 0.5 $\mu$m for the highest-quality sample TC5 at 0 T---a value more than three orders of magnitude larger than the lattice constant $a$ $\sim$ 3.9 \AA, supporting the high mobility of low-energy spin excitations.

Along the SSL contour, the group velocity $v_\mathrm{m0}$ $\sim$ $\nabla_{\mathbf{Q}_\mathrm{G}}\omega$ (where $\hbar\omega$ is the excitation energy) is expected to be small due to ground-state degeneracy. However, weak quantum fluctuations from $S$ = 7/2 and perturbations beyond the classical easy-plane frustrated honeycomb model may slightly lift this degeneracy, resulting in a nonzero $v_\mathrm{m0}$ in GdZnPO. This likely explains why the measured $\lambda_\mathrm{m0}v_\mathrm{m0}$ is nonzero.

\begin{figure}[h]
\begin{center}
\includegraphics[width=9cm,angle=0]{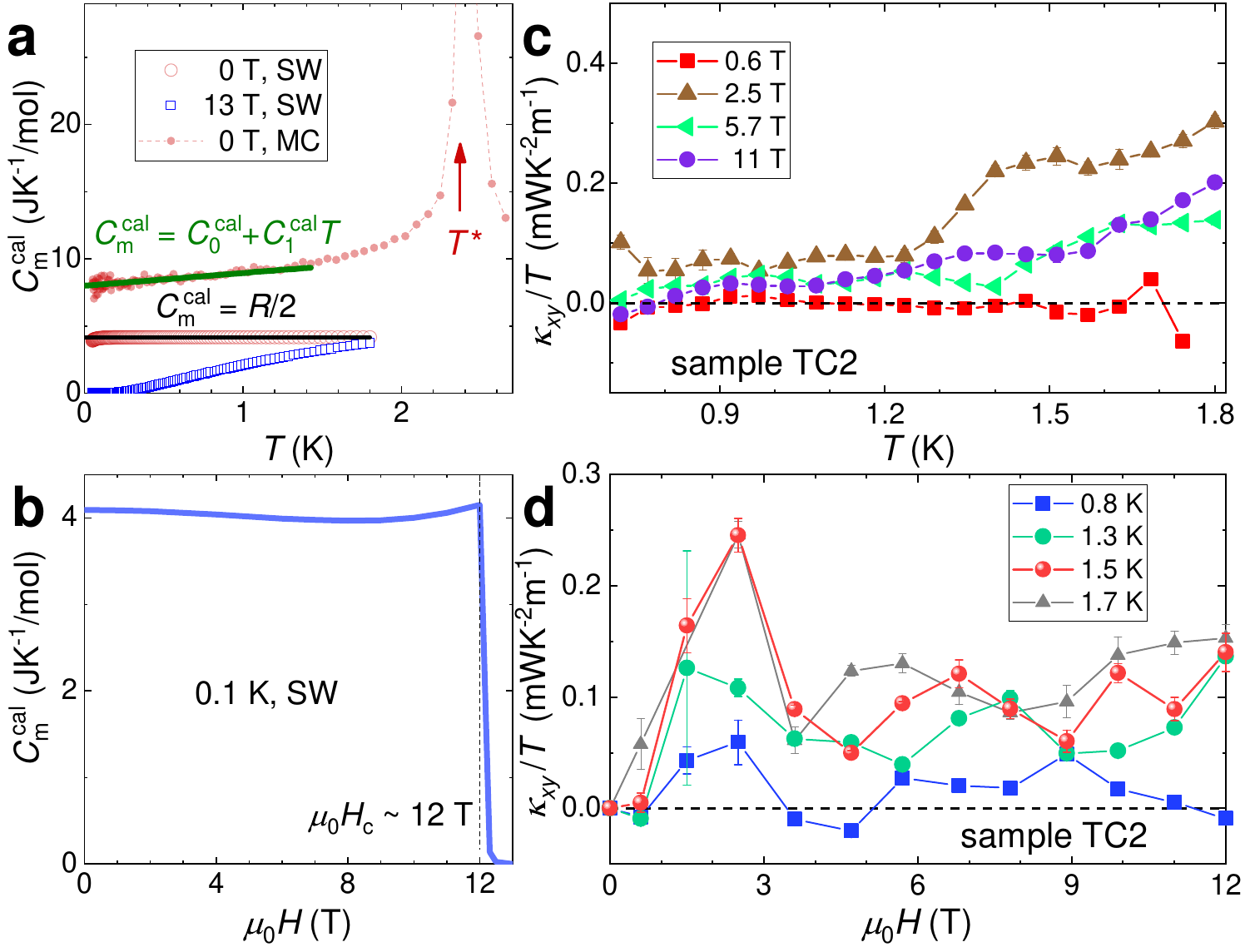}
\captionsetup{labelformat=empty,font={small}}
\caption{\textbf{Fig. 4 $\vert$ Calculated specific heat and thermal Hall conductivity ($\kappa_{xy}$) of GdZnPO sample TC2.}
\textbf{a}, Specific heat ($C_\mathrm{m}^\mathrm{cal}$) calculated using spin-wave (SW) theory and Monte Carlo (MC) simulations. The black line represents the $C_\mathrm{m}^\mathrm{cal}$ = $R$/2 behavior, while the olive line depicts a linear fit to the MC data using $C_\mathrm{m}^\mathrm{cal}$ = $C_0^\mathrm{cal}$+$C_1^\mathrm{cal}T$. The crossover temperature $T^*$ is indicated. \textbf{b}, Field dependence of $C_\mathrm{m}^\mathrm{cal}$ at 0.1 K. The dashed line marks the crossover field $\mu_0H_\mathrm{c}$ $\sim$ 12 T. \textbf{c}, Temperature dependence of $\kappa_{xy}$/$T$ at selected magnetic fields. \textbf{d}, Magnetic field dependence of $\kappa_{xy}$/$T$ measured at selected temperatures. In \textbf{c},\textbf{d}, error bars, 1$\sigma$ s.e.}
\label{fig4}
\end{center}
\end{figure}

At low temperatures, $\kappa_{xx}$/$T$ decreases with increasing field and exhibits a broad hump around 6 T, while $C_\mathrm{m}$/$T$ remains nearly constant up to $\sim$11 T  before showing a sharp drop near the crossover field $\mu_0H_\mathrm{c}$ $\sim$ 12 T in agreement with spin-wave calculations (see Figs.~\ref{fig3}c,d and \ref{fig4}b). Since the low-temperature ($\sim$0.1 K) heat transport is dominated by spin excitations along the spiral contour, the observed hump in $\kappa_{xx}$/$T$ near 6 T and the drop in $C_\mathrm{m}$/$T$ near $\mu_0H_\mathrm{c}$ are mainly governed by the residual terms $\kappa_0$ and $C_0$, respectively (Fig.~\ref{fig3}e,f). Within a quasiparticle framework, the distinct field dependencies of $\kappa_0$ and $C_0$ arise from variations in the product $\lambda_\mathrm{m0}v_\mathrm{m0}$, which is generally field-dependent. Even at $\mu_0H_\mathrm{c}$ $\sim$ 12 T, the GdZnPO spin system is not fully polarized, as indicated by the measured susceptibility d$M$/d$H$ remaining larger than the Van Vleck value ($\chi_\mathrm{vv}^\parallel$ = 0.3 cm$^3$/mol)~\cite{PhysRevLett.133.236704} at 50 mK (Supplementary Fig. 13), likely due to quantum fluctuations associated with the finite spin quantum number $S$ = 7/2. As a result, both $\kappa_0$ and $C_0$ are significantly suppressed but remain finite at 12 T (see Fig.~\ref{fig3}e,f and Supplementary Fig. 13c,d). The absence of a sharp drop in $\kappa_0$ suggests a sudden enhancement in the mean free path and/or group velocity of the excitations along the spiral contour, i.e., in $\lambda_\mathrm{m0}v_\mathrm{m0}$, likely driven by quantum fluctuations near $\mu_0H_\mathrm{c}$ (see Supplementary Fig. 13e). In contrast, the linear term $\kappa_1$ (Fig.~\ref{fig3}e), and thus $\lambda_\mathrm{m1}v_\mathrm{m1}$, gradually decreases with increasing field up to 12 T, in rough agreement with the field dependence of the mean magnon velocity calculated for excitations off the SSL contour (see Supplementary Fig. 18c). While current models of magnetic thermal transport remain limited, the temperature and field dependence of $C_\mathrm{m}$ is qualitatively captured by spin-wave theory and Monte Carlo simulations (Fig.~\ref{fig4}a,b).

\bigskip

\noindent {\bf Thermal Hall effect}

\noindent The thermal Hall effect serves as a powerful probe of spin excitations~\cite{onose2010observation,hirschberger2015large,PhysRevLett.104.066403,kasahara2018majorana,PhysRevLett.124.186602,czajka2023planar}. Figure~\ref{fig4}c presents the thermal Hall conductivity, $\kappa_{xy}$/$T$, measured on the high-quality GdZnPO sample TC2 with a magnetic field applied along the $c$ axis. At $\sim$1.5 K---well below the representative phonon Debye temperature $\Theta_\mathrm{D}$ $\sim$ 117.5 K---a clear positive $\kappa_{xy}$/$T$ is observed (Fig.~\ref{fig4}c), and is expected to significantly exceed the phonon contribution~\cite{PhysRevLett.124.105901}. Given that the Zeeman energy associated with $\sim$12 T is only $\sim$10 K, much smaller than the phonon energy scale ($>$100 K), a linear field dependence of the phonon thermal Hall conductivity, $\kappa_{xy}^\mathrm{p}$ $\sim$ $a\mu_0H$, is generally expected below 12 T within the linear response regime, where $a$ is a constant~\cite{PhysRevLett.95.155901,PhysRevLett.118.145902,PhysRevLett.124.105901}. Through a linear fit to the measured $\kappa_{xy}$ between $\sim$6 and 12 T, the estimated phonon contribution, $a\mu_0H$, remains significantly smaller than $\kappa_{xy}$ across the entire field range (see Supplementary Fig. 15c). Since GdZnPO is an insulator, the observed $\kappa_{xy}$/$T$ is expected to predominantly arise from spin excitations. The residual value of observed $\kappa_{xy}/T$ approaches zero below $\sim$0.8 K (Fig.~\ref{fig4}c), supporting a dominate bosonic origin~\cite{PhysRevLett.124.186602,czajka2023planar}. On the honeycomb lattice, each unit cell contains two spins (see Fig.~\ref{fig1}b), resulting in two bands. The temperature dependence of $\kappa_{xy}$/$T$ can be approximated by the two-flat-band bosonic model~\cite{czajka2023planar}, with the lower band's Chern number $C_s$ $\sim$ 1 (Supplementary Fig. 15a), supporting the topological nature of (at least some) spin excitations in GdZnPO.\\

As shown in Fig.~\ref{fig4}d, $\kappa_{xy}$/$T$ exhibits a peak at $\sim$ 2.5 T, a feature strikingly similar to that recently observed in the SSL candidate MnSc$_2$S$_4$ at comparably low temperatures~\cite{takeda2024magnon}. In that system, the peak has been attributed to a magnon thermal Hall effect arising from antiferromagnetic skyrmions at high fields. In GdZnPO, the classical easy-plane frustrated honeycomb model also supports spin and momentum vortices with nonzero winding numbers at finite fields~\cite{PhysRevLett.133.236704}, which may contribute to the observed thermal Hall effect (see Fig.~\ref{fig4}c,d). Possible origins are further discussed in the next section. An oscillation-like feature in $\kappa_{xy}$/$T$ appears at 1.5 K (see Fig.~\ref{fig4}d), but no clear periodicity in $\mu_0H$, $\mu_0^{-1}H^{-1}$~\cite{zhang2024large}, or log$_{10}$($\mu_0H$) is identified (see Supplementary Fig. 15). In addition, the pattern of $\kappa_{xy}$/$T$ changes noticeably above $\sim$4 T at a different temperature (see Fig.~\ref{fig4}d), further suggesting that the apparent oscillations likely stem from ordinary measurement noise.

\bigskip

\section*{Discussion}

\begin{figure}[h]
\begin{center}
\includegraphics[width=9cm,angle=0]{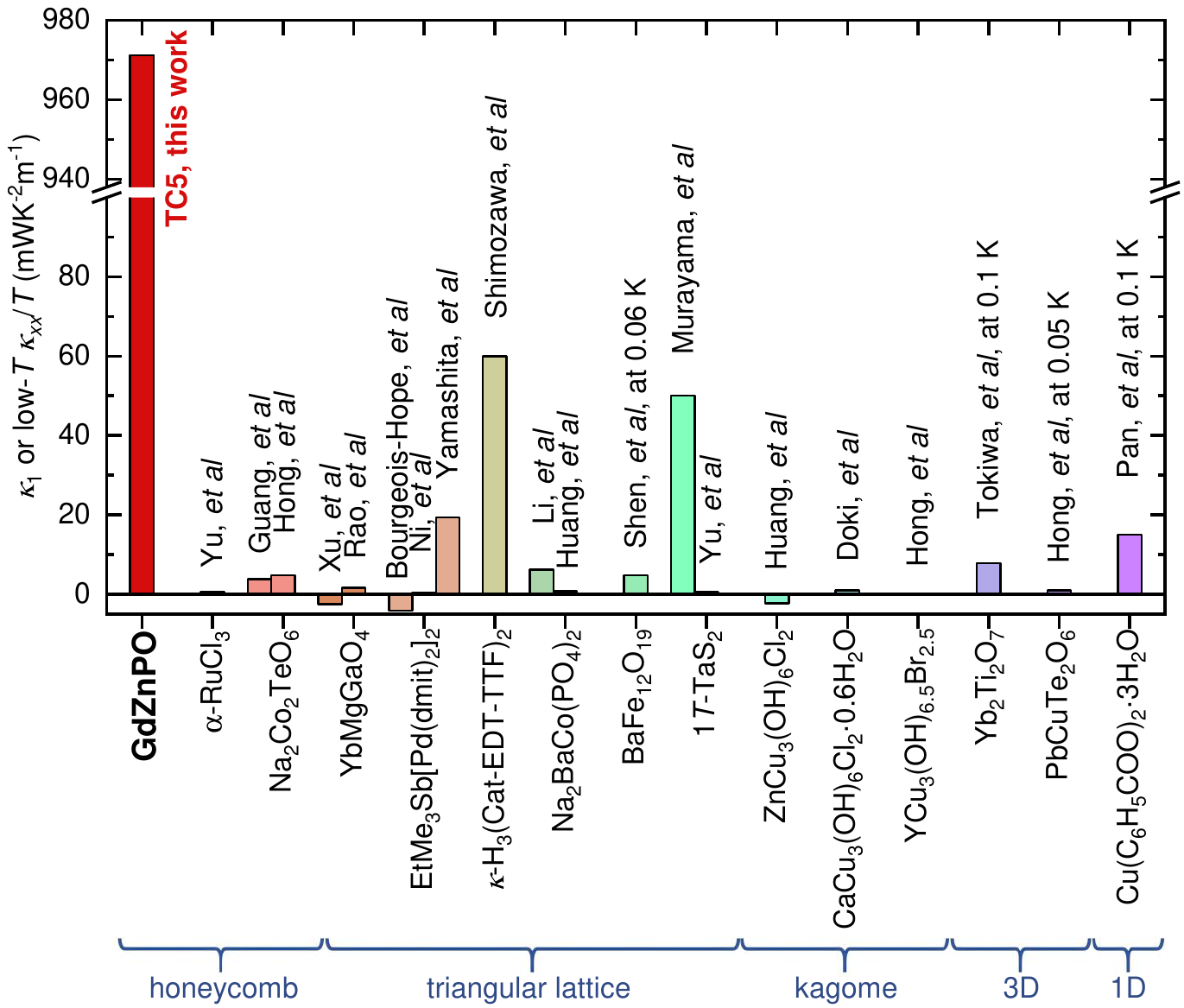}
\captionsetup{labelformat=empty,font={small}}
\caption{\textbf{Fig. 5 $\vert$ Low-temperature thermal conductivity, $\kappa_1$, of GdZnPO sample TC5 compared to other magnetic insulators.}
Previously reported $\kappa_1$ or low-temperature $\kappa_{xx}$/$T$ values are shown for various magnetic insulators, including two-dimensional honeycomb $\alpha$-RuCl$_3$~\cite{PhysRevLett.120.067202} and Na$_2$Co$_2$TeO$_6$~\cite{PhysRevB.107.184423,hong2024phonon}; two-dimensional triangular-lattice YbMgGaO$_4$~\cite{PhysRevLett.117.267202,rao2021survival}, EtMe$_3$Sb[Pd(dmit)$_2$]$_2$~\cite{PhysRevX.9.041051,PhysRevLett.123.247204,PhysRevB.101.140407}, $\kappa$-H$_3$(Cat-EDT-TTF)$_2$~\cite{shimozawa2017quantum}, Na$_2$BaCo(PO$_4$)$_2$~\cite{li2020possible,huang2022thermal}, BaFe$_{12}$O$_{19}$~\cite{shen2016quantum}, and 1$T$-TaS$_2$~\cite{PhysRevResearch.2.013099,PhysRevB.96.081111}; two-dimensional kagome ZnCu$_3$(OH)$_6$Cl$_2$~\cite{PhysRevLett.127.267202}, CaCu$_3$(OH)$_6$Cl$_2\cdot$0.6H$_2$O~\cite{PhysRevLett.121.097203}, and YCu$_{3}$(OH)$_{6.5}$Br$_{2.5}$~\cite{PhysRevB.106.L220406}; three-dimensional (3D) Yb$_2$Ti$_2$O$_7$~\cite{tokiwa2016possible} and PbCuTe$_{2}$O$_{6}$~\cite{PhysRevLett.131.256701}; and one-dimensional (1D) antiferromagnetic Cu(C$_6$H$_5$COO)$_2\cdot$3H$_2$O~\cite{PhysRevLett.129.167201}.}
\label{fig5}
\end{center}
\end{figure}

The existence of nonzero magnetic thermal conductivity in the low-$T$ limit, $\kappa_1$, remains a debated topic among various magnetic insulators, including prominent spin-liquid candidates, as shown in Fig.~\ref{fig5}~\cite{PhysRevLett.120.067202,PhysRevB.107.184423,hong2024phonon,PhysRevLett.117.267202,rao2021survival,10.1126/science.1188200,PhysRevX.9.041051,PhysRevLett.123.247204,PhysRevB.101.140407,shimozawa2017quantum,li2020possible,huang2022thermal,PhysRevB.104.104403,shen2016quantum,PhysRevLett.127.267202,PhysRevLett.121.097203,PhysRevB.106.L220406,jeon2024one,tokiwa2016possible,PhysRevLett.131.256701,PhysRevLett.129.167201,PhysRevResearch.2.013099,PhysRevB.96.081111}. The triangular-lattice spin-liquid candidate 1$T$-TaS$_2$ achieves a high reported value of $\kappa_1$ $\leq$ 50 mWK$^{-2}$m$^{-1}$~\cite{PhysRevResearch.2.013099}, though this remains a topic of debate~\cite{PhysRevB.96.081111}. Notably, 1$T$-TaS$_{2}$ exhibits a resistivity below 1 $\Omega$m down to 0.5 K~\cite{PhysRevResearch.2.013099}, over three orders of magnitude smaller than that of GdZnPO (Fig.~\ref{fig2}c). Similarly, the spin-liquid candidate $\kappa$-H$_3$(Cat-EDT-TTF)$_2$ exhibits a comparably high value of $\kappa_1$ $\sim$ 60 mWK$^{-2}$m$^{-1}$~\cite{shimozawa2017quantum}. Consequently, GdZnPO emerges as a rare magnetic insulator with remarkably high intrinsic magnetic thermal conductivity, following $\kappa_{xx}^\mathrm{m}$ $\sim$ $\kappa_1T$+$\kappa_0$ in the low-$T$ limit, with $\kappa_1$ $\sim$ 970 mWK$^{-2}$m$^{-1}$ (see Fig.~\ref{fig3}b) and $\kappa_0$ $\sim$ 250 mWK$^{-1}$m$^{-1}$ (see Fig.~\ref{fig3}a) at $\sim$ 0 T from the highest-quality crystal TC5. The observed thermal conductivity in the low-$T$ limit, $\kappa_1$, is exceptionally large compared to the other magnetic insulators (see Fig.~\ref{fig5}). More importantly, the substantial residual thermal conductivity $\kappa_0$ (at $\sim$0 T), attributed to spin excitations along the putative SSL contour in GdZnPO, has not been reported in other magnetic insulators to the best of our knowledge. These findings suggest the presence of intrinsic, mobile, high-density low-energy spin excitations and the stability of the putative SSL, without order by disorder, down to at least $\sim$0.05 K in GdZnPO.

Because of the strong neutron absorption by Gd atoms, neutron scattering measurements on GdZnPO remain challenging. Nevertheless, several experimental observations support the emergence of a SSL in GdZnPO at low temperatures: (1) Low-temperature magnetization measurements clearly reveal easy-plane anisotropy. Combined with the crystal structure and magnetization data, the spin system of GdZnPO is well described by the $S$ = 7/2 easy-plane $J_1$-$J_2$ honeycomb-lattice model~\cite{PhysRevLett.133.236704}. Theoretically, a SSL is stabilized within this model across a broad parameter range, $\mid$$J_2/J_1$$\mid$ $>$ 1/6, persisting down to very low temperatures~\cite{PhysRevResearch.4.013121}. Furthermore, the experimentally determined Hamiltonian yields a crossover field of $\mu_0H_\mathrm{c}$ $\sim$ 12 T and a Curie-Weiss temperature of $\theta_\mathrm{w}$ = $-S(S+1)(J_1+2J_2)$ $\sim$ $-$12 K, both of which are in good agreement with the experimental results on GdZnPO (see Fig.~\ref{fig3}d,f)~\cite{PhysRevLett.133.236704}. (2) Within the spherical approximation, the generic SSL theory predicts a low-temperature magnetic specific heat of $C_\mathrm{m}$ $\sim$ $C_0+C_1T$~\cite{yao2021generic}---a distinctive behavior not reported in other spin systems to our knowledge. Experimentally, GdZnPO exhibits this feature at low temperatures up to the crossover field $\mu_0H_\mathrm{c}$ $\sim$ 12 T, see Fig.~\ref{fig1}c. (3) Within the SSL ansatz on the honeycomb lattice, the zero-temperature susceptibility is theoretically predicted to be constant up to $\sim\mu_0H_\mathrm{c}$: $\chi_\mathrm{cal}^{\parallel}$ = $\mu_0N_\mathrm{A}g^2\mu_\mathrm{B}^2/[2D+3J_1+9J_2+J_1^2/(4J_2)]$+$\chi_\mathrm{vv}^\parallel$ ($\sim$4.4 cm$^3$/mol), consistent with the experimental observations down to 50 mK ($\sim$0.4\%$\mid$$\theta_\mathrm{w}$$\mid$, see Supplementary Figs. 13a and 14a). (4) The giant magnetocaloric effect observed in GdZnPO is well explained by the determined easy-plane $J_1$-$J_2$ honeycomb-lattice spin Hamiltonian without tuning parameters~\cite{zhao2025giant}. (5) As shown in Fig.~\ref{fig3}a,b, the temperature dependence of thermal conductivity closely follows the expected form $\kappa_{xx}$ = $\kappa_0+\kappa_1T+K_\mathrm{p}T^3$ below 1 K and up to the crossover field $\mu_0H_\mathrm{c}$ $\sim$ 12 T. Within the quasi-particle framework, the magnetic thermal conductivity is proportional to the magnetic specific heat, i.e., $\kappa_{xx}^\mathrm{m}$ ($\sim$ $\kappa_0+\kappa_1T$) $\propto$ $C_\mathrm{m}$. Notably, the observation of a magnetic contribution $\kappa_{xx}^\mathrm{m}$ $\sim$ $\kappa_0+\kappa_1T$ is highly distinctive and, to our knowledge, has not been reported in any other magnetic compound. This unusual thermal transport behavior further substantiates the emergence of a SSL in GdZnPO.

In the ordered phases of the pure easy-plane frustrated honeycomb model, linear spin-wave bands possess zero Chern numbers, implying no thermal Hall effect~\cite{PhysRevB.106.035113}. Thus, the positive thermal Hall effect observed in GdZnPO may stem from several factors: (1) The calculations~\cite{PhysRevB.106.035113} assume perfect SSL ground-state configurations, while low-energy topological excitations, like spin and momentum vortices, emerge at low temperatures~\cite{PhysRevLett.133.236704}, potentially yielding nonzero Chern numbers for the magnon bands via a fictitious magnetic flux~\cite{mascot2021topological}. These topological excitations may also contribute directly to the thermal Hall effect above $\sim$1.2 K (Fig.~\ref{fig4}c) through a momentum-transfer force~\cite{PhysRevB.90.094423} if they are mobile. (2) Weak perturbations, such as symmetrically allowed (Fig.~\ref{fig1}a) DM interactions, may induce nonzero Chern numbers and a magnon thermal Hall effect~\cite{PhysRevB.106.035113}. (3) The low-energy structure of the spiral contour may accommodate other thermal excitations, that contribute to thermal Hall transport. (4) While the calculations are classical~\cite{PhysRevB.106.035113}, GdZnPO's $S$ = 7/2 system exhibits weak quantum fluctuations.

The putative 2D SSL in the frustrated honeycomb-lattice antiferromagnet GdZnPO has been shown to persist from $T^* \sim 2$ K down to $\sim$0.053 K, confirming its stability with a high density of low-energy spin excitations at $H < H_c$~\cite{PhysRevLett.133.236704}. This work further demonstrates that these excitations are not only mobile but also at least partially topological, as evidenced by the observed giant low-temperature thermal conductivity (down to $\sim$0.05 K) and positive thermal Hall effect. These findings provide new avenues for exploring the exotic transport and topological properties of low-lying excitations in SSL candidates.

\section*{Methods}

GdZnPO single crystals were grown using the flux method~\cite{PhysRevLett.133.236704}. Despite nearly identical synthesis procedures, the crystal sizes, Laue FWHMs, electric resistivities, and low-temperature thermal conductivities vary significantly between samples (see Fig.~\ref{fig2}). In contrast, the low-temperature specific heat shows weak sample dependence across all three measured samples, as shown in Supplementary Fig. 10. The as-grown oxide single crystals of GdZnPO are very thin (thickness $\lesssim$ 0.1 mm) and mechanically fragile. High-resolution measurements of both thermal conductivity and specific heat require relatively large crystals---with length $\gtrsim$ 1 mm or mass $\gtrsim$ 1 mg. Unfortunately, crystals often broke into smaller fragments during the careful removal of silver paste (used in thermal conductivity measurements) or GE varnish (used in specific heat measurements), making it impractical at present to measure both properties on the same GdZnPO crystal. Sample TC3 was confirmed to have broken during the thermal-conductivity experiment at fields above $\sim$6 T due to the torque induced by the easy-plane anisotropy of the spin system, and thus we failed to collect its higher-field data.

The as-grown crystals' longest dimension aligns with the [110] direction, as determined by Laue XRD, and thermal or electrical currents were applied along this direction (Fig.~\ref{fig1}b). Longitudinal thermal conductivity and thermal Hall conductivity were measured using standard four- and five-wire steady-state methods~\cite{zhao2024quantum}, respectively, in magnetic fields up to 12 T along the $c$ axis and temperatures ranging from 0.05 to 2 K, achieved with a superconducting magnet and a $^3$He-$^4$He dilution refrigerator (Supplementary Note 1 and Supplementary Figs. 1-12). We employed a cantilever-based thermal conductivity setup that was thermally isolated from all supporting components, with only minor additional heat flow occurring primarily through the NbTi superconducting leads (13 mm in length and 63 $\mu$m in diameter) connected to the heater and thermometers (Supplementary Fig. 7c). The resulting heat-leakage conductance is negligibly small, as detailed in Supplementary Note 1.

Specific heat down to 43 mK and magnetization down to 50 mK were also measured in the $^3$He-$^4$He dilution refrigerator~\cite{PhysRevLett.133.236704}. Crystal quality was accessed via the FWHM of Laue XRD reflections measured under identical conditions. Electrical resistance measurements were conducted using a PPMS (Quantum Design). Linear spin-wave theory was employed to simulate the specific heat and mean magnon velocity using the previously determined GdZnPO Hamiltonian~\cite{PhysRevLett.133.236704}, without parameter tuning (Supplementary Note 3 and Supplementary Figs. 17 and 18). Standard MC simulations of the specific heat were performed on a 2$\times$60$^2$ cluster with periodic boundary conditions, using the same Hamiltonian. Each simulation comprised 15,000 MC steps at each of 200 temperatures, annealing gradually from 50 K to 0.05 K, with 5,000 steps allocated for thermalization.

\section*{Data availability}

The data generated within the main text are provided in the Source Data. Additional raw data are available from the corresponding authors upon request.

\backmatter
\section*{Acknowledgments}
We gratefully acknowledge Xiaokang Li, Shang Gao, Zhengxin Liu, Changle Liu, and Haijun Liao for helpful discussion. This work was supported by the National Key R\&D Program of China (Grant Nos. 2024YFA1613100 (Y.L.) and 2023YFA1406500 (Y.L.)), the National Natural Science Foundation of China (No. 12274153 (Y.L.)), and the Fundamental Research Funds for the Central Universities (No. HUST: 2020kfyXJJS054 (Y.L.)).

\section*{Author contributions}
Y.L. planed and supervised the project. Y.Z. and Y.L. collected the thermal transport, specific heat, Laue x-ray diffraction, and electrical resistance data. X.C., Z.W., and Z.M. prepared the GdZnPO single crystals. Y.L. and X.C. conducted the spin-wave calculations and Monte Carlo simulations. Y.L., Y.Z., X.Y., and X.H. analysed the data and wrote the manuscript with comments from all co-authors. The manuscript reflects the contributions of all authors.

\section*{Competing interests}
The authors declare no competing interests.

\section*{Additional information}

\noindent{\bf Supplementary Information}
This file contains Supplementary Notes 1-3, Supplementary Figs. 1-18, Supplementary Table 1, and References.

\section*{Figure Legends}

\bigskip

\bigskip
\bigskip

\bigskip
\bigskip

\bigskip
\bigskip

\bigskip
\bigskip

\bigskip






\begin{thebibliography}{10}
\expandafter\ifx\csname url\endcsname\relax
  \def\url#1{\burl{#1}}\fi
\expandafter\ifx\csname urlprefix\endcsname\relax\def\urlprefix{URL }\fi
\providecommand{\bibinfo}[2]{#2}
\providecommand{\eprint}[2][]{\url{#2}}
\providecommand{\doi}[1]{\url{https://doi.org/#1}}
\bibcommenthead

\bibitem{savary2016quantum}
\bibinfo{author}{Savary, L.} \& \bibinfo{author}{Balents, L.}
\newblock \bibinfo{title}{Quantum spin liquids: a review}.
\newblock \emph{\bibinfo{journal}{Rep. Prog. Phy.}}
  \textbf{\bibinfo{volume}{80}}, \bibinfo{pages}{016502}
  (\bibinfo{year}{2016}).
\newblock
  \urlprefix\url{https://iopscience.iop.org/article/10.1088/0034-4885/80/1/016502/meta}
  .

\bibitem{zhou2017quantum}
\bibinfo{author}{Zhou, Y.}, \bibinfo{author}{Kanoda, K.} \&
  \bibinfo{author}{Ng, T.-K.}
\newblock \bibinfo{title}{Quantum spin liquid states}.
\newblock \emph{\bibinfo{journal}{Rev. Mod. Phys.}}
  \textbf{\bibinfo{volume}{89}}, \bibinfo{pages}{025003}
  (\bibinfo{year}{2017}).
\newblock
  \urlprefix\url{https://journals.aps.org/rmp/abstract/10.1103/RevModPhys.89.025003}
  .

\bibitem{nayak2008non}
\bibinfo{author}{Nayak, C.}, \bibinfo{author}{Simon, S.~H.},
  \bibinfo{author}{Stern, A.}, \bibinfo{author}{Freedman, M.} \&
  \bibinfo{author}{Das~Sarma, S.}
\newblock \bibinfo{title}{{Non-Abelian} anyons and topological quantum
  computation}.
\newblock \emph{\bibinfo{journal}{Rev. Mod. Phys.}}
  \textbf{\bibinfo{volume}{80}}, \bibinfo{pages}{1083} (\bibinfo{year}{2008}).
\newblock
  \urlprefix\url{https://journals.aps.org/rmp/abstract/10.1103/RevModPhys.80.1083}
  .

\bibitem{jungwirth2016antiferromagnetic}
\bibinfo{author}{Jungwirth, T.}, \bibinfo{author}{Marti, X.},
  \bibinfo{author}{Wadley, P.} \& \bibinfo{author}{Wunderlich, J.}
\newblock \bibinfo{title}{Antiferromagnetic spintronics}.
\newblock \emph{\bibinfo{journal}{Nat. Nanotechnol.}}
  \textbf{\bibinfo{volume}{11}}, \bibinfo{pages}{231} (\bibinfo{year}{2016}).
\newblock \urlprefix\url{https://www.nature.com/articles/nnano.2016.18} .

\bibitem{gao2020fractional}
\bibinfo{author}{Gao, S.} \emph{et~al.}
\newblock \bibinfo{title}{Fractional antiferromagnetic skyrmion lattice induced
  by anisotropic couplings}.
\newblock \emph{\bibinfo{journal}{Nature}} \textbf{\bibinfo{volume}{586}},
  \bibinfo{pages}{37} (\bibinfo{year}{2020}).
\newblock \urlprefix\url{https://doi.org/10.1038/s41586-020-2716-8} .

\bibitem{RevModPhys.78.17}
\bibinfo{author}{Lee, P.~A.}, \bibinfo{author}{Nagaosa, N.} \&
  \bibinfo{author}{Wen, X.-G.}
\newblock \bibinfo{title}{Doping a {Mott} insulator: Physics of
  high-temperature superconductivity}.
\newblock \emph{\bibinfo{journal}{Rev. Mod. Phys.}}
  \textbf{\bibinfo{volume}{78}}, \bibinfo{pages}{17} (\bibinfo{year}{2006}).
\newblock \urlprefix\url{https://link.aps.org/doi/10.1103/RevModPhys.78.17} .

\bibitem{bergman2007order}
\bibinfo{author}{Bergman, D.}, \bibinfo{author}{Alicea, J.},
  \bibinfo{author}{Gull, E.}, \bibinfo{author}{Trebst, S.} \&
  \bibinfo{author}{Balents, L.}
\newblock \bibinfo{title}{Order-by-disorder and spiral spin-liquid in
  frustrated diamond-lattice antiferromagnets}.
\newblock \emph{\bibinfo{journal}{Nat. Phys.}} \textbf{\bibinfo{volume}{3}},
  \bibinfo{pages}{487} (\bibinfo{year}{2007}).
\newblock \urlprefix\url{https://www.nature.com/articles/nphys622} .

\bibitem{yao2021generic}
\bibinfo{author}{Yao, X.-P.}, \bibinfo{author}{Liu, J.~Q.},
  \bibinfo{author}{Huang, C.-J.}, \bibinfo{author}{Wang, X.} \&
  \bibinfo{author}{Chen, G.}
\newblock \bibinfo{title}{Generic spiral spin liquids}.
\newblock \emph{\bibinfo{journal}{Front. Phys.}} \textbf{\bibinfo{volume}{16}},
  \bibinfo{pages}{53303} (\bibinfo{year}{2021}).
\newblock
  \urlprefix\url{https://link.springer.com/article/10.1007/s11467-021-1074-9} .

\bibitem{PhysRevB.81.214419}
\bibinfo{author}{Mulder, A.}, \bibinfo{author}{Ganesh, R.},
  \bibinfo{author}{Capriotti, L.} \& \bibinfo{author}{Paramekanti, A.}
\newblock \bibinfo{title}{Spiral order by disorder and lattice nematic order in
  a frustrated {Heisenberg} antiferromagnet on the honeycomb lattice}.
\newblock \emph{\bibinfo{journal}{Phys. Rev. B}} \textbf{\bibinfo{volume}{81}},
  \bibinfo{pages}{214419} (\bibinfo{year}{2010}).
\newblock \urlprefix\url{https://link.aps.org/doi/10.1103/PhysRevB.81.214419} .

\bibitem{owerre2017topological}
\bibinfo{author}{Owerre, S.~A.}
\newblock \bibinfo{title}{Topological magnon bands and unconventional thermal
  {Hall} effect on the frustrated honeycomb and bilayer triangular lattice}.
\newblock \emph{\bibinfo{journal}{J. Phys.: Condens. Matter}}
  \textbf{\bibinfo{volume}{29}}, \bibinfo{pages}{385801}
  (\bibinfo{year}{2017}).
\newblock
  \urlprefix\url{https://iopscience.iop.org/article/10.1088/1361-648X/aa7dd2/meta}
  .

\bibitem{PhysRevB.106.035113}
\bibinfo{author}{Fujiwara, K.}, \bibinfo{author}{Kitamura, S.} \&
  \bibinfo{author}{Morimoto, T.}
\newblock \bibinfo{title}{Thermal {Hall} responses in frustrated honeycomb spin
  systems}.
\newblock \emph{\bibinfo{journal}{Phys. Rev. B}}
  \textbf{\bibinfo{volume}{106}}, \bibinfo{pages}{035113}
  (\bibinfo{year}{2022}).
\newblock \urlprefix\url{https://link.aps.org/doi/10.1103/PhysRevB.106.035113}
  .

\bibitem{okumura2010novel}
\bibinfo{author}{Okumura, S.}, \bibinfo{author}{Kawamura, H.},
  \bibinfo{author}{Okubo, T.} \& \bibinfo{author}{Motome, Y.}
\newblock \bibinfo{title}{Novel spin-liquid states in the frustrated
  {Heisenberg} antiferromagnet on the honeycomb lattice}.
\newblock \emph{\bibinfo{journal}{J. Phys. Soc. Jpn.}}
  \textbf{\bibinfo{volume}{79}}, \bibinfo{pages}{114705}
  (\bibinfo{year}{2010}).
\newblock \urlprefix\url{https://journals.jps.jp/doi/10.1143/JPSJ.79.114705} .

\bibitem{PhysRevB.100.224404}
\bibinfo{author}{Shimokawa, T.}, \bibinfo{author}{Okubo, T.} \&
  \bibinfo{author}{Kawamura, H.}
\newblock \bibinfo{title}{Multiple-$q$ states of the
  ${J}_{1}\ensuremath{-}{J}_{2}$ classical honeycomb-lattice {Heisenberg}
  antiferromagnet under a magnetic field}.
\newblock \emph{\bibinfo{journal}{Phys. Rev. B}}
  \textbf{\bibinfo{volume}{100}}, \bibinfo{pages}{224404}
  (\bibinfo{year}{2019}).
\newblock \urlprefix\url{https://link.aps.org/doi/10.1103/PhysRevB.100.224404}
  .

\bibitem{PhysRevResearch.4.013121}
\bibinfo{author}{Huang, C.-J.}, \bibinfo{author}{Liu, J.~Q.} \&
  \bibinfo{author}{Chen, G.}
\newblock \bibinfo{title}{Spiral spin liquid behavior and persistent reciprocal
  kagome structure in frustrated van der {Waals} magnets and beyond}.
\newblock \emph{\bibinfo{journal}{Phys. Rev. Res.}}
  \textbf{\bibinfo{volume}{4}}, \bibinfo{pages}{013121} (\bibinfo{year}{2022}).
\newblock
  \urlprefix\url{https://link.aps.org/doi/10.1103/PhysRevResearch.4.013121} .

\bibitem{PhysRevLett.133.236704}
\bibinfo{author}{Wan, Z.} \emph{et~al.}
\newblock \bibinfo{title}{Spiral spin liquid in a frustrated honeycomb
  antiferromagnet: A single-crystal study of {GdZnPO}}.
\newblock \emph{\bibinfo{journal}{Phys. Rev. Lett.}}
  \textbf{\bibinfo{volume}{133}}, \bibinfo{pages}{236704}
  (\bibinfo{year}{2024}).
\newblock
  \urlprefix\url{https://link.aps.org/doi/10.1103/PhysRevLett.133.236704} .

\bibitem{yao2013topologically}
\bibinfo{author}{Yao, N.~Y.} \emph{et~al.}
\newblock \bibinfo{title}{Topologically protected quantum state transfer in a
  chiral spin liquid}.
\newblock \emph{\bibinfo{journal}{Nat. Commun.}} \textbf{\bibinfo{volume}{4}},
  \bibinfo{pages}{1585} (\bibinfo{year}{2013}).
\newblock \urlprefix\url{https://www.nature.com/articles/ncomms2531} .

\bibitem{PhysRevB.95.115139}
\bibinfo{author}{Pretko, M.}
\newblock \bibinfo{title}{Subdimensional particle structure of higher rank
  {$U(1)$} spin liquids}.
\newblock \emph{\bibinfo{journal}{Phys. Rev. B}} \textbf{\bibinfo{volume}{95}},
  \bibinfo{pages}{115139} (\bibinfo{year}{2017}).
\newblock \urlprefix\url{https://link.aps.org/doi/10.1103/PhysRevB.95.115139} .

\bibitem{nandkishore2019fractons}
\bibinfo{author}{Nandkishore, R.~M.} \& \bibinfo{author}{Hermele, M.}
\newblock \bibinfo{title}{Fractons}.
\newblock \emph{\bibinfo{journal}{Annu. Rev. Condens. Matter Phys.}}
  \textbf{\bibinfo{volume}{10}}, \bibinfo{pages}{295} (\bibinfo{year}{2019}).
\newblock
  \urlprefix\url{https://doi.org/10.1146/annurev-conmatphys-031218-013604} .

\bibitem{pretko2020fracton}
\bibinfo{author}{Pretko, M.}, \bibinfo{author}{Chen, X.} \&
  \bibinfo{author}{You, Y.}
\newblock \bibinfo{title}{Fracton phases of matter}.
\newblock \emph{\bibinfo{journal}{Int. J. Mod. Phys. A}}
  \textbf{\bibinfo{volume}{35}}~(06), \bibinfo{pages}{2030003}
  (\bibinfo{year}{2020}).
\newblock \urlprefix\url{https://doi.org/10.1142/S0217751X20300033} .

\bibitem{PhysRevResearch.4.023175}
\bibinfo{author}{Yan, H.} \& \bibinfo{author}{Reuther, J.}
\newblock \bibinfo{title}{Low-energy structure of spiral spin liquids}.
\newblock \emph{\bibinfo{journal}{Phys. Rev. Res.}}
  \textbf{\bibinfo{volume}{4}}, \bibinfo{pages}{023175} (\bibinfo{year}{2022}).
\newblock
  \urlprefix\url{https://link.aps.org/doi/10.1103/PhysRevResearch.4.023175} .

\bibitem{PhysRevB.108.224423}
\bibinfo{author}{Hsieh, T.-C.} \& \bibinfo{author}{Radzihovsky, L.}
\newblock \bibinfo{title}{{$O(N)$} smectic $\ensuremath{\sigma}$ model}.
\newblock \emph{\bibinfo{journal}{Phys. Rev. B}}
  \textbf{\bibinfo{volume}{108}}, \bibinfo{pages}{224423}
  (\bibinfo{year}{2023}).
\newblock \urlprefix\url{https://link.aps.org/doi/10.1103/PhysRevB.108.224423}
  .

\bibitem{zhao2025giant}
\bibinfo{author}{Zhao, Y.}, \bibinfo{author}{Chen, X.}, \bibinfo{author}{Wan,
  Z.}, \bibinfo{author}{Ma, Z.} \& \bibinfo{author}{Li, Y.}
\newblock \bibinfo{title}{Giant magnetocaloric effect in a honeycomb-lattice
  spiral spin liquid candidate}.
\newblock \emph{\bibinfo{journal}{arXiv:2508.08721}}  (\bibinfo{year}{2025,
  accepted in Advanced Science}) .

\bibitem{PhysRevLett.120.067202}
\bibinfo{author}{Yu, Y.~J.} \emph{et~al.}
\newblock \bibinfo{title}{Ultralow-temperature thermal conductivity of the
  {Kitaev} honeycomb magnet {$\alpha$-RuCl$_3$} across the field-induced phase
  transition}.
\newblock \emph{\bibinfo{journal}{Phys. Rev. Lett.}}
  \textbf{\bibinfo{volume}{120}}, \bibinfo{pages}{067202}
  (\bibinfo{year}{2018}).
\newblock
  \urlprefix\url{https://link.aps.org/doi/10.1103/PhysRevLett.120.067202} .

\bibitem{PhysRevB.107.184423}
\bibinfo{author}{Guang, S.} \emph{et~al.}
\newblock \bibinfo{title}{Thermal transport of fractionalized antiferromagnetic
  and field-induced states in the {Kitaev} material
  {Na$_{2}$Co$_{2}$TeO$_{6}$}}.
\newblock \emph{\bibinfo{journal}{Phys. Rev. B}}
  \textbf{\bibinfo{volume}{107}}, \bibinfo{pages}{184423}
  (\bibinfo{year}{2023}).
\newblock \urlprefix\url{https://link.aps.org/doi/10.1103/PhysRevB.107.184423}
  .

\bibitem{hong2024phonon}
\bibinfo{author}{Hong, X.} \emph{et~al.}
\newblock \bibinfo{title}{Phonon thermal transport shaped by strong spin-phonon
  scattering in a {Kitaev} material {Na$_{2}$Co$_{2}$TeO$_{6}$}}.
\newblock \emph{\bibinfo{journal}{npj Quantum Mater.}}
  \textbf{\bibinfo{volume}{9}}, \bibinfo{pages}{18} (\bibinfo{year}{2024}).
\newblock \urlprefix\url{https://www.nature.com/articles/s41535-024-00628-4} .

\bibitem{PhysRevLett.117.267202}
\bibinfo{author}{Xu, Y.} \emph{et~al.}
\newblock \bibinfo{title}{Absence of magnetic thermal conductivity in the
  quantum spin-liquid candidate {YbMgGaO$_4$}}.
\newblock \emph{\bibinfo{journal}{Phys. Rev. Lett.}}
  \textbf{\bibinfo{volume}{117}}, \bibinfo{pages}{267202}
  (\bibinfo{year}{2016}).
\newblock
  \urlprefix\url{https://link.aps.org/doi/10.1103/PhysRevLett.117.267202} .

\bibitem{rao2021survival}
\bibinfo{author}{Rao, X.} \emph{et~al.}
\newblock \bibinfo{title}{Survival of itinerant excitations and quantum spin
  state transitions in {YbMgGaO$_4$} with chemical disorder}.
\newblock \emph{\bibinfo{journal}{Nat. Commun.}} \textbf{\bibinfo{volume}{12}},
  \bibinfo{pages}{4949} (\bibinfo{year}{2021}).
\newblock \urlprefix\url{https://www.nature.com/articles/s41467-021-25247-6} .

\bibitem{10.1126/science.1188200}
\bibinfo{author}{Yamashita, M.} \emph{et~al.}
\newblock \bibinfo{title}{Highly mobile gapless excitations in a
  two-dimensional candidate quantum spin liquid}.
\newblock \emph{\bibinfo{journal}{Science}} \textbf{\bibinfo{volume}{328}},
  \bibinfo{pages}{1246} (\bibinfo{year}{2010}).
\newblock
  \urlprefix\url{https://www.science.org/doi/abs/10.1126/science.1188200} .

\bibitem{PhysRevX.9.041051}
\bibinfo{author}{Bourgeois-Hope, P.} \emph{et~al.}
\newblock \bibinfo{title}{Thermal conductivity of the quantum spin liquid
  candidate {EtMe$_{3}$Sb[Pd(dmit)$_{2}$]$_{2}$}: No evidence of mobile gapless
  excitations}.
\newblock \emph{\bibinfo{journal}{Phys. Rev. X}} \textbf{\bibinfo{volume}{9}},
  \bibinfo{pages}{041051} (\bibinfo{year}{2019}).
\newblock \urlprefix\url{https://link.aps.org/doi/10.1103/PhysRevX.9.041051} .

\bibitem{PhysRevLett.123.247204}
\bibinfo{author}{Ni, J.~M.} \emph{et~al.}
\newblock \bibinfo{title}{Absence of magnetic thermal conductivity in the
  quantum spin liquid candidate {EtMe$_{3}$Sb[Pd(dmit)$_{2}$]$_{2}$}}.
\newblock \emph{\bibinfo{journal}{Phys. Rev. Lett.}}
  \textbf{\bibinfo{volume}{123}}, \bibinfo{pages}{247204}
  (\bibinfo{year}{2019}).
\newblock
  \urlprefix\url{https://link.aps.org/doi/10.1103/PhysRevLett.123.247204} .

\bibitem{PhysRevB.101.140407}
\bibinfo{author}{Yamashita, M.} \emph{et~al.}
\newblock \bibinfo{title}{Presence and absence of itinerant gapless excitations
  in the quantum spin liquid candidate {EtMe$_{3}$Sb[Pd(dmit)$_{2}$]$_{2}$}}.
\newblock \emph{\bibinfo{journal}{Phys. Rev. B}}
  \textbf{\bibinfo{volume}{101}}, \bibinfo{pages}{140407}
  (\bibinfo{year}{2020}).
\newblock \urlprefix\url{https://link.aps.org/doi/10.1103/PhysRevB.101.140407}
  .

\bibitem{shimozawa2017quantum}
\bibinfo{author}{Shimozawa, M.} \emph{et~al.}
\newblock \bibinfo{title}{Quantum-disordered state of magnetic and electric
  dipoles in an organic {Mott} system}.
\newblock \emph{\bibinfo{journal}{Nat. Commun.}} \textbf{\bibinfo{volume}{8}},
  \bibinfo{pages}{1821} (\bibinfo{year}{2017}).
\newblock \urlprefix\url{https://doi.org/10.1038/s41467-017-01849-x} .

\bibitem{li2020possible}
\bibinfo{author}{Li, N.} \emph{et~al.}
\newblock \bibinfo{title}{Possible itinerant excitations and quantum spin state
  transitions in the effective spin-1/2 triangular-lattice antiferromagnet
  {Na$_2$BaCo(PO$_4$)$_2$}}.
\newblock \emph{\bibinfo{journal}{Nat. Commun.}} \textbf{\bibinfo{volume}{11}},
  \bibinfo{pages}{4216} (\bibinfo{year}{2020}).
\newblock \urlprefix\url{https://doi.org/10.1038/s41467-020-18041-3} .

\bibitem{huang2022thermal}
\bibinfo{author}{Huang, Y.~Y.} \emph{et~al.}
\newblock \bibinfo{title}{Thermal conductivity of triangular-lattice
  antiferromagnet {Na$_2$BaCo(PO$_4$)$_2$}: Absence of itinerant fermionic
  excitations}  (\bibinfo{year}{2022}).
\newblock \urlprefix\url{https://arxiv.org/abs/2206.08866} .

\bibitem{PhysRevB.104.104403}
\bibinfo{author}{Li, N.} \emph{et~al.}
\newblock \bibinfo{title}{Quantum spin state transitions in the spin-1
  equilateral triangular lattice antiferromagnet
  {Na$_{2}$BaNi(PO$_{4}$)$_{2}$}}.
\newblock \emph{\bibinfo{journal}{Phys. Rev. B}}
  \textbf{\bibinfo{volume}{104}}, \bibinfo{pages}{104403}
  (\bibinfo{year}{2021}).
\newblock \urlprefix\url{https://link.aps.org/doi/10.1103/PhysRevB.104.104403}
  .

\bibitem{shen2016quantum}
\bibinfo{author}{Shen, S.-P.} \emph{et~al.}
\newblock \bibinfo{title}{Quantum electric-dipole liquid on a triangular
  lattice}.
\newblock \emph{\bibinfo{journal}{Nat. Commun.}} \textbf{\bibinfo{volume}{7}},
  \bibinfo{pages}{10569} (\bibinfo{year}{2016}).
\newblock \urlprefix\url{https://doi.org/10.1038/ncomms10569} .

\bibitem{PhysRevLett.127.267202}
\bibinfo{author}{Huang, Y.~Y.} \emph{et~al.}
\newblock \bibinfo{title}{Heat transport in herbertsmithite: Can a quantum spin
  liquid survive disorder?}
\newblock \emph{\bibinfo{journal}{Phys. Rev. Lett.}}
  \textbf{\bibinfo{volume}{127}}, \bibinfo{pages}{267202}
  (\bibinfo{year}{2021}).
\newblock
  \urlprefix\url{https://link.aps.org/doi/10.1103/PhysRevLett.127.267202} .

\bibitem{PhysRevLett.121.097203}
\bibinfo{author}{Doki, H.} \emph{et~al.}
\newblock \bibinfo{title}{Spin thermal {Hall} conductivity of a kagome
  antiferromagnet}.
\newblock \emph{\bibinfo{journal}{Phys. Rev. Lett.}}
  \textbf{\bibinfo{volume}{121}}, \bibinfo{pages}{097203}
  (\bibinfo{year}{2018}).
\newblock
  \urlprefix\url{https://link.aps.org/doi/10.1103/PhysRevLett.121.097203} .

\bibitem{PhysRevB.106.L220406}
\bibinfo{author}{Hong, X.} \emph{et~al.}
\newblock \bibinfo{title}{Heat transport of the kagome {Heisenberg} quantum
  spin liquid candidate {YCu$_{3}$(OH)$_{6.5}$Br$_{2.5}$}: Localized magnetic
  excitations and a putative spin gap}.
\newblock \emph{\bibinfo{journal}{Phys. Rev. B}}
  \textbf{\bibinfo{volume}{106}}, \bibinfo{pages}{L220406}
  (\bibinfo{year}{2022}).
\newblock \urlprefix\url{https://link.aps.org/doi/10.1103/PhysRevB.106.L220406}
  .

\bibitem{jeon2024one}
\bibinfo{author}{Jeon, S.} \emph{et~al.}
\newblock \bibinfo{title}{One-ninth magnetization plateau stabilized by spin
  entanglement in a kagome antiferromagnet}.
\newblock \emph{\bibinfo{journal}{Nat. Phys.}} \textbf{\bibinfo{volume}{20}},
  \bibinfo{pages}{435} (\bibinfo{year}{2024}).
\newblock \urlprefix\url{https://doi.org/10.1038/s41567-023-02318-7} .

\bibitem{tokiwa2016possible}
\bibinfo{author}{Tokiwa, Y.} \emph{et~al.}
\newblock \bibinfo{title}{Possible observation of highly itinerant quantum
  magnetic monopoles in the frustrated pyrochlore {Yb$_2$Ti$_2$O$_7$}}.
\newblock \emph{\bibinfo{journal}{Nat. Commun.}} \textbf{\bibinfo{volume}{7}},
  \bibinfo{pages}{10807} (\bibinfo{year}{2016}).
\newblock \urlprefix\url{https://doi.org/10.1038/ncomms10807} .

\bibitem{PhysRevLett.131.256701}
\bibinfo{author}{Hong, X.} \emph{et~al.}
\newblock \bibinfo{title}{Spinon heat transport in the three-dimensional
  quantum magnet {PbCuTe$_{2}$O$_{6}$}}.
\newblock \emph{\bibinfo{journal}{Phys. Rev. Lett.}}
  \textbf{\bibinfo{volume}{131}}, \bibinfo{pages}{256701}
  (\bibinfo{year}{2023}).
\newblock
  \urlprefix\url{https://link.aps.org/doi/10.1103/PhysRevLett.131.256701} .

\bibitem{PhysRevLett.129.167201}
\bibinfo{author}{Pan, B.~Y.} \emph{et~al.}
\newblock \bibinfo{title}{Unambiguous experimental verification of
  linear-in-temperature spinon thermal conductivity in an antiferromagnetic
  {Heisenberg} chain}.
\newblock \emph{\bibinfo{journal}{Phys. Rev. Lett.}}
  \textbf{\bibinfo{volume}{129}}, \bibinfo{pages}{167201}
  (\bibinfo{year}{2022}).
\newblock
  \urlprefix\url{https://link.aps.org/doi/10.1103/PhysRevLett.129.167201} .

\bibitem{PhysRevResearch.2.013099}
\bibinfo{author}{Murayama, H.} \emph{et~al.}
\newblock \bibinfo{title}{Effect of quenched disorder on the quantum spin
  liquid state of the triangular-lattice antiferromagnet {1$T$-TaS$_{2}$}}.
\newblock \emph{\bibinfo{journal}{Phys. Rev. Res.}}
  \textbf{\bibinfo{volume}{2}}, \bibinfo{pages}{013099} (\bibinfo{year}{2020}).
\newblock
  \urlprefix\url{https://link.aps.org/doi/10.1103/PhysRevResearch.2.013099} .

\bibitem{PhysRevB.96.081111}
\bibinfo{author}{Yu, Y.~J.} \emph{et~al.}
\newblock \bibinfo{title}{Heat transport study of the spin liquid candidate
  {1$T$-TaS$_{2}$}}.
\newblock \emph{\bibinfo{journal}{Phys. Rev. B}} \textbf{\bibinfo{volume}{96}},
  \bibinfo{pages}{081111} (\bibinfo{year}{2017}).
\newblock \urlprefix\url{https://link.aps.org/doi/10.1103/PhysRevB.96.081111} .

\bibitem{nientiedt1998equiatomic}
\bibinfo{author}{Nientiedt, A.~T.} \& \bibinfo{author}{Jeitschko, W.}
\newblock \bibinfo{title}{Equiatomic quaternary rare earth element zinc
  pnictide oxides {RZnPO} and {RZnAsO}}.
\newblock \emph{\bibinfo{journal}{Inorg. Chem.}} \textbf{\bibinfo{volume}{37}},
  \bibinfo{pages}{386} (\bibinfo{year}{1998}).
\newblock \urlprefix\url{https://pubs.acs.org/doi/full/10.1021/ic971058q} .

\bibitem{lincke2008magnetic}
\bibinfo{author}{Lincke, H.} \emph{et~al.}
\newblock \bibinfo{title}{Magnetic, optical, and electronic properties of the
  phosphide oxides {REZnPO} ({RE} = {Y}, {La}-{Nd}, {Sm}, {Gd}, {Dy}, {Ho})}.
\newblock \emph{\bibinfo{journal}{Z. Anorg. Allg. Chem.}}
  \textbf{\bibinfo{volume}{634}}, \bibinfo{pages}{1339} (\bibinfo{year}{2008}).
\newblock \urlprefix\url{https://doi.org/10.1002/zaac.200800066} .

\bibitem{PhysRev.120.91}
\bibinfo{author}{Moriya, T.}
\newblock \bibinfo{title}{Anisotropic superexchange interaction and weak
  ferromagnetism}.
\newblock \emph{\bibinfo{journal}{Phys. Rev.}} \textbf{\bibinfo{volume}{120}},
  \bibinfo{pages}{91} (\bibinfo{year}{1960}).
\newblock \urlprefix\url{https://link.aps.org/doi/10.1103/PhysRev.120.91} .

\bibitem{PhysRevB.86.045314}
\bibinfo{author}{Xiong, J.} \emph{et~al.}
\newblock \bibinfo{title}{High-field {Shubnikov}--de {Haas} oscillations in the
  topological insulator {Bi$_2$Te$_2$Se}}.
\newblock \emph{\bibinfo{journal}{Phys. Rev. B}} \textbf{\bibinfo{volume}{86}},
  \bibinfo{pages}{045314} (\bibinfo{year}{2012}).
\newblock \urlprefix\url{https://link.aps.org/doi/10.1103/PhysRevB.86.045314} .

\bibitem{PhysRevB.105.024418}
\bibinfo{author}{Liu, J.} \emph{et~al.}
\newblock \bibinfo{title}{Gapless spin liquid behavior in a kagome {Heisenberg}
  antiferromagnet with randomly distributed hexagons of alternate bonds}.
\newblock \emph{\bibinfo{journal}{Phys. Rev. B}}
  \textbf{\bibinfo{volume}{105}}, \bibinfo{pages}{024418}
  (\bibinfo{year}{2022}).
\newblock \urlprefix\url{https://link.aps.org/doi/10.1103/PhysRevB.105.024418}
  .

\bibitem{Liu2021Frustrated}
\bibinfo{author}{Liu, J.} \emph{et~al.}
\newblock \bibinfo{title}{Frustrated magnetism of the triangular-lattice
  antiferromagnets {$\alpha$-CrOOH} and {$\alpha$-CrOOD}}.
\newblock \emph{\bibinfo{journal}{New J. Phys.}} \textbf{\bibinfo{volume}{23}},
  \bibinfo{pages}{033040} (\bibinfo{year}{2021}).
\newblock \urlprefix\url{https://doi.org/10.1088/1367-2630/abe813} .

\bibitem{gofryk2014anisotropic}
\bibinfo{author}{Gofryk, K.} \emph{et~al.}
\newblock \bibinfo{title}{Anisotropic thermal conductivity in uranium dioxide}.
\newblock \emph{\bibinfo{journal}{Nat. Commun.}} \textbf{\bibinfo{volume}{5}},
  \bibinfo{pages}{4551} (\bibinfo{year}{2014}).
\newblock \urlprefix\url{https://www.nature.com/articles/ncomms5551} .

\bibitem{PhysRevB.104.144426}
\bibinfo{author}{Hong, X.} \emph{et~al.}
\newblock \bibinfo{title}{Strongly scattered phonon heat transport of the
  candidate {Kitaev} material {Na$_2$Co$_2$TeO$_6$}}.
\newblock \emph{\bibinfo{journal}{Phys. Rev. B}}
  \textbf{\bibinfo{volume}{104}}, \bibinfo{pages}{144426}
  (\bibinfo{year}{2021}).
\newblock \urlprefix\url{https://link.aps.org/doi/10.1103/PhysRevB.104.144426}
  .

\bibitem{zhao2024quantum}
\bibinfo{author}{Zhao, Y.} \emph{et~al.}
\newblock \bibinfo{title}{Quantum annealing of a frustrated magnet}.
\newblock \emph{\bibinfo{journal}{Nat. Commun.}} \textbf{\bibinfo{volume}{15}},
  \bibinfo{pages}{3495} (\bibinfo{year}{2024}).
\newblock \urlprefix\url{https://doi.org/10.1038/s41467-024-47819-y} .

\bibitem{PhysRevLett.110.217209}
\bibinfo{author}{Toews, W.~H.} \emph{et~al.}
\newblock \bibinfo{title}{Thermal conductivity of {Ho$_{2}$Ti$_{2}$O$_{7}$}
  along the [111] direction}.
\newblock \emph{\bibinfo{journal}{Phys. Rev. Lett.}}
  \textbf{\bibinfo{volume}{110}}, \bibinfo{pages}{217209}
  (\bibinfo{year}{2013}).
\newblock
  \urlprefix\url{https://link.aps.org/doi/10.1103/PhysRevLett.110.217209} .

\bibitem{li2015gapless}
\bibinfo{author}{Li, Y.} \emph{et~al.}
\newblock \bibinfo{title}{Gapless quantum spin liquid ground state in the
  two-dimensional spin-1/2 triangular antiferromagnet {YbMgGaO$_4$}}.
\newblock \emph{\bibinfo{journal}{Sci. Rep.}} \textbf{\bibinfo{volume}{5}},
  \bibinfo{pages}{16419} (\bibinfo{year}{2015}).
\newblock \urlprefix\url{https://doi.org/10.1038/srep16419} .

\bibitem{paddison2017continuous}
\bibinfo{author}{Paddison, J. A.~M.} \emph{et~al.}
\newblock \bibinfo{title}{Continuous excitations of the triangular-lattice
  quantum spin liquid {YbMgGaO$_4$}}.
\newblock \emph{\bibinfo{journal}{Nat. Phys.}} \textbf{\bibinfo{volume}{13}},
  \bibinfo{pages}{117} (\bibinfo{year}{2017}).
\newblock \urlprefix\url{https://doi.org/10.1038/nphys3971} .

\bibitem{onose2010observation}
\bibinfo{author}{Onose, Y.} \emph{et~al.}
\newblock \bibinfo{title}{Observation of the magnon {Hall} effect}.
\newblock \emph{\bibinfo{journal}{Science}} \textbf{\bibinfo{volume}{329}},
  \bibinfo{pages}{297} (\bibinfo{year}{2010}).
\newblock
  \urlprefix\url{https://www.science.org/doi/abs/10.1126/science.1188260} .

\bibitem{hirschberger2015large}
\bibinfo{author}{Hirschberger, M.}, \bibinfo{author}{Krizan, J.~W.},
  \bibinfo{author}{Cava, R.~J.} \& \bibinfo{author}{Ong, N.~P.}
\newblock \bibinfo{title}{Large thermal {Hall} conductivity of neutral spin
  excitations in a frustrated quantum magnet}.
\newblock \emph{\bibinfo{journal}{Science}} \textbf{\bibinfo{volume}{348}},
  \bibinfo{pages}{106} (\bibinfo{year}{2015}).
\newblock
  \urlprefix\url{https://www.science.org/doi/abs/10.1126/science.1257340} .

\bibitem{PhysRevLett.104.066403}
\bibinfo{author}{Katsura, H.}, \bibinfo{author}{Nagaosa, N.} \&
  \bibinfo{author}{Lee, P.~A.}
\newblock \bibinfo{title}{Theory of the thermal {Hall} effect in quantum
  magnets}.
\newblock \emph{\bibinfo{journal}{Phys. Rev. Lett.}}
  \textbf{\bibinfo{volume}{104}}, \bibinfo{pages}{066403}
  (\bibinfo{year}{2010}).
\newblock
  \urlprefix\url{https://link.aps.org/doi/10.1103/PhysRevLett.104.066403} .

\bibitem{kasahara2018majorana}
\bibinfo{author}{Kasahara, Y.} \emph{et~al.}
\newblock \bibinfo{title}{Majorana quantization and half-integer thermal
  quantum {Hall} effect in a {Kitaev} spin liquid}.
\newblock \emph{\bibinfo{journal}{Nature}} \textbf{\bibinfo{volume}{559}},
  \bibinfo{pages}{227} (\bibinfo{year}{2018}).
\newblock \urlprefix\url{https://doi.org/10.1038/s41586-018-0274-0} .

\bibitem{PhysRevLett.124.186602}
\bibinfo{author}{Yang, Y.}, \bibinfo{author}{Zhang, G.} \&
  \bibinfo{author}{Zhang, F.}
\newblock \bibinfo{title}{Universal behavior of the thermal {Hall}
  conductivity}.
\newblock \emph{\bibinfo{journal}{Phys. Rev. Lett.}}
  \textbf{\bibinfo{volume}{124}}, \bibinfo{pages}{186602}
  (\bibinfo{year}{2020}).
\newblock
  \urlprefix\url{https://link.aps.org/doi/10.1103/PhysRevLett.124.186602}.
\newblock \doi{10.1103/PhysRevLett.124.186602} .

\bibitem{czajka2023planar}
\bibinfo{author}{Czajka, P.} \emph{et~al.}
\newblock \bibinfo{title}{Planar thermal {Hall} effect of topological bosons in
  the {Kitaev} magnet $\alpha$-{RuCl$_3$}}.
\newblock \emph{\bibinfo{journal}{Nat. Mater.}} \textbf{\bibinfo{volume}{22}},
  \bibinfo{pages}{36} (\bibinfo{year}{2023}).
\newblock \urlprefix\url{https://doi.org/10.1038/s41563-022-01397-w} .

\bibitem{PhysRevLett.124.105901}
\bibinfo{author}{Li, X.}, \bibinfo{author}{Fauqu{\'e}, B.},
  \bibinfo{author}{Zhu, Z.} \& \bibinfo{author}{Behnia, K.}
\newblock \bibinfo{title}{Phonon thermal {Hall} effect in strontium titanate}.
\newblock \emph{\bibinfo{journal}{Phys. Rev. Lett.}}
  \textbf{\bibinfo{volume}{124}}, \bibinfo{pages}{105901}
  (\bibinfo{year}{2020}).
\newblock
  \urlprefix\url{https://link.aps.org/doi/10.1103/PhysRevLett.124.105901} .

\bibitem{PhysRevLett.95.155901}
\bibinfo{author}{Strohm, C.}, \bibinfo{author}{Rikken, G. L. J.~A.} \&
  \bibinfo{author}{Wyder, P.}
\newblock \bibinfo{title}{Phenomenological evidence for the phonon {Hall}
  effect}.
\newblock \emph{\bibinfo{journal}{Phys. Rev. Lett.}}
  \textbf{\bibinfo{volume}{95}}, \bibinfo{pages}{155901}
  (\bibinfo{year}{2005}).
\newblock
  \urlprefix\url{https://link.aps.org/doi/10.1103/PhysRevLett.95.155901} .

\bibitem{PhysRevLett.118.145902}
\bibinfo{author}{Sugii, K.} \emph{et~al.}
\newblock \bibinfo{title}{Thermal {Hall} effect in a phonon-glass
  {Ba$_3$CuSb$_2$O$_9$}}.
\newblock \emph{\bibinfo{journal}{Phys. Rev. Lett.}}
  \textbf{\bibinfo{volume}{118}}, \bibinfo{pages}{145902}
  (\bibinfo{year}{2017}).
\newblock
  \urlprefix\url{https://link.aps.org/doi/10.1103/PhysRevLett.118.145902} .

\bibitem{takeda2024magnon}
\bibinfo{author}{Takeda, H.} \emph{et~al.}
\newblock \bibinfo{title}{Magnon thermal {Hall} effect via emergent {SU}(3)
  flux on the antiferromagnetic skyrmion lattice}.
\newblock \emph{\bibinfo{journal}{Nat. Commun.}} \textbf{\bibinfo{volume}{15}},
  \bibinfo{pages}{566} (\bibinfo{year}{2024}).
\newblock \urlprefix\url{https://www.nature.com/articles/s41467-024-44793-3} .

\bibitem{zhang2024large}
\bibinfo{author}{Zhang, D.} \emph{et~al.}
\newblock \bibinfo{title}{Large oscillatory thermal hall effect in kagome
  metals}.
\newblock \emph{\bibinfo{journal}{Nat. Commun.}} \textbf{\bibinfo{volume}{15}},
  \bibinfo{pages}{6224} (\bibinfo{year}{2024}).
\newblock \urlprefix\url{https://www.nature.com/articles/s41467-024-50336-7} .

\bibitem{mascot2021topological}
\bibinfo{author}{Mascot, E.}, \bibinfo{author}{Bedow, J.},
  \bibinfo{author}{Graham, M.}, \bibinfo{author}{Rachel, S.} \&
  \bibinfo{author}{Morr, D.~K.}
\newblock \bibinfo{title}{Topological superconductivity in skyrmion lattices}.
\newblock \emph{\bibinfo{journal}{npj Quantum Mater.}}
  \textbf{\bibinfo{volume}{6}}, \bibinfo{pages}{6} (\bibinfo{year}{2021}).
\newblock \urlprefix\url{https://www.nature.com/articles/s41535-020-00299-x} .

\bibitem{PhysRevB.90.094423}
\bibinfo{author}{Sch\"utte, C.} \& \bibinfo{author}{Garst, M.}
\newblock \bibinfo{title}{Magnon-skyrmion scattering in chiral magnets}.
\newblock \emph{\bibinfo{journal}{Phys. Rev. B}} \textbf{\bibinfo{volume}{90}},
  \bibinfo{pages}{094423} (\bibinfo{year}{2014}).
\newblock \urlprefix\url{https://link.aps.org/doi/10.1103/PhysRevB.90.094423} .

\end{thebibliography}
\end{document}